\documentclass[aps,prl,twocolumn,amsmath,amssymb,floatfix,longbibliography,superscriptaddress]{revtex4-2}
\usepackage{color}
\usepackage{mathrsfs}
\usepackage{graphicx}
\usepackage{dcolumn}
\usepackage{bm}
\usepackage{empheq}
\usepackage[hidelinks]{hyperref}
\usepackage{xcolor}
\hypersetup{
    colorlinks,
    linkcolor={blue},
    citecolor={blue!80!black},
    urlcolor={blue!80!black}
}
\usepackage[toc,page]{appendix}
\usepackage[normalem]{ulem}
\usepackage{amsmath,amsfonts,amssymb,ulem}
\usepackage{epstopdf}
\usepackage{xcolor}
\usepackage[separate-uncertainty]{siunitx}
\usepackage{upgreek}
\usepackage{hyperref} % Optional for \autoref

\newcommand{\re}[1]{{\color{black}#1}}

\usepackage{graphicx}
\makeatletter
\def\@fnsymbol#1{\ensuremath{\ifcase#1\or a)\or b)\or c)\or d)\else\@ctrerr\fi}}
\makeatother

\begin{document}

\title{Wavefront-Guided Electron Injection for Direct Laser Acceleration in Relativistic Laser-Driven Plasma Channel}

\author{Heng Wang}
\affiliation{Department of Engineering Physics, Tsinghua University, Beijing 100084, China}
\author{Liang Sheng}
% \email[]{shengliang@nint.ac.cn}
\thanks{Authors to whom correspondence should be addressed: shengliang@nint.ac.cn, zgong92@itp.ac.cn}
\affiliation{National Key Laboratory of Intense Pulsed Radiation Simulation and Effect, Northwest Institute of Nuclear Technology, Xi'an 710024, China}
\author{Long Zeng}
\affiliation{Department of Engineering Physics, Tsinghua University, Beijing 100084, China}
\author{Yang Li}
\affiliation{National Key Laboratory of Intense Pulsed Radiation Simulation and Effect, Northwest Institute of Nuclear Technology, Xi'an 710024, China}
\author{Zheng Gong}
% \email[]{zgong92@itp.ac.cn}
\thanks{Authors to whom correspondence should be addressed: shengliang@nint.ac.cn, zgong92@itp.ac.cn}
\affiliation{Institute of Theoretical Physics, Chinese Academy of Sciences, Beijing 100190, China}

\date{\today}
\begin{abstract}

We investigate electron injection into direct laser acceleration (DLA) in relativistic laser-driven plasma channels using particle-in-cell simulations. We identify and characterize a wavefront-guided injection mechanism, in which electrons are continuously fed into the plasma channel through the density pile-up layer at the \re{laser-pulse} front. Phase-space analysis reveals a localized injectable region within the pile-up layer, indicating that only a selected subset of electrons satisfies the conditions required for the subsequent direct laser acceleration. This mechanism provides a physical interpretation for the high-charge capability of DLA by explaining how electrons are continuously supplied to the accelerating channel. Beyond this continuous supply process, the injection dynamics are further modulated by \re{the periodic variation of the carrier phase at the laser-pulse front.} The spatial locations of injected electrons are found to be closely associated with magnetic-island structures formed under laser-phase modulation, suggesting that the laser wavefront not only supplies electrons but also organizes their entry into the accelerating channel. These findings advance the physical understanding of energetic-electron generation in relativistic laser-driven subcritical-density plasma channels and are relevant to the development of compact DLA-based particle and radiation sources.

\end{abstract}
\maketitle

\section{Introduction}
In plasmas, direct laser acceleration (DLA) provides an important route to generating high-charge energetic electron beams. Compared with conventional radio-frequency accelerators, laser-plasma accelerators can sustain extremely large accelerating fields over microscopic distances, making them a promising platform for compact high-energy electron sources \cite{Tajima1979, Leemans2014,Pukhov2003}. Among the different laser-plasma acceleration schemes, laser wakefield acceleration (LWFA) has achieved GeV and even multi-GeV electron beams, but it generally operates in relatively low-density plasmas and produces electron bunches with comparatively limited charge \cite{Leemans2006, Esarey2009, Pukhov2003}. DLA, by contrast, can operate in higher-density plasmas and laser-driven plasma channels, where electrons can gain energy continuously from the laser field while undergoing transverse motion in self-generated channel fields\cite{Pukhov1999, Gahn1999, Mangles2005, Arefiev2012, Arefiev2016Beyond, Arefiev2016Nonplanar, Khudik2016, Gong2020}. This capability makes DLA particularly attractive for producing electron beams that combine high charge with high energy \cite{Pukhov1999, Gahn1999, Mangles2005, Hussein2021, Babjak2024, Tavana2026}. Early theoretical and experimental studies showed that, in relativistic laser-driven plasma channels, electrons can directly extract energy from the laser field through a resonance between their transverse betatron oscillations and the Doppler-shifted laser field \cite{Pukhov1999, Gahn1999,Arefiev2012}. Subsequent petawatt-class laser experiments further established DLA as a key mechanism for generating high-charge energetic electron beams \cite{Mangles2005, Hussein2021}.

DLA has attracted broad interest not only as an efficient electron energization mechanism but also as a source of high-current electron beams for compact radiation and particle sources. Its importance for secondary-source development lies in the fact that DLA-driven electrons can act as a primary energy carrier linking intense laser pulses to multiple downstream conversion processes. In plasma channels and near-critical-density targets, these electrons can generate bright betatron-like x-ray and gamma-ray emission \cite{Stark2016, huang2016characteristics, gong2018brilliant, Rosmej2021, Cikhardt2024,Gyrdymov2024,Meir2025,Tangtartharakul2025}, whereas their interaction with high-Z converters can produce bremsstrahlung photons that drive photonuclear reactions and neutron generation \cite{Rosmej2020, Tavana2026,Gunther2022,Cohen2024,Tavana2023, Cohen2026Stabilizing}. Related electron-driven schemes have further been explored for positron production \cite{Mathiak2026, Martinez2025PRE, Martinez2023PRAB, InigoGamiz2025}, isotope generation \cite{Rosmej2020}, and coherent electromagnetic emission \cite{Gorlova2024}, as well as THz transition radiation from laser-accelerated electron bunches. In structured or relativistically transparent targets, DLA-enhanced electron transport can also strengthen sheath-field formation and target charging, thereby promoting efficient proton and ion acceleration \cite{Bin2015, Horny2022, wang2021super}.
These results indicate that DLA is not only a fundamental channel of energy transfer in intense laser-plasma interactions, but also an important physical basis for developing bright radiation sources and high-flux particle sources. 

However, compared with the acceleration stage of DLA \cite{Arefiev2012, Arefiev2014, Arefiev2016Beyond, Khudik2016, Arefiev2016Nonplanar, Robinson2020, Arefiev2020, Gong2019, Gong2020, Wang2020, Arefiev2024,Yeh2025}, the injection process that precedes electron trapping into the accelerating phase remains far less understood. Electron injection is a central issue in laser-plasma interactions because it determines which electrons can enter the acceleration region dominated by the intense laser field and, consequently, affects the charge, energy spectrum, divergence, and spatiotemporal structure of the resulting electron beam \cite{Esarey2009}. In this sense, injection is not merely an initial condition for subsequent acceleration. Rather, it connects laser propagation, plasma-channel formation, self-generated electromagnetic-field evolution, and electron energization into a single physical chain. It also provides an important source of variation among different experimental observations and acceleration regimes. Understanding and controlling electron injection are therefore essential for producing high-quality electron beams and for improving the performance of secondary sources driven by these beams, including bright x-ray and gamma-ray radiation, energetic ions, positrons, muons, and polarized particles\cite{Hussein2021,Tang2025POP, Reichwein2025Plasma}. This capability is also central to the broader development of compact laser-plasma accelerators. 

Previous studies have discussed DLA electron injection in several specific contexts. In hybrid LWFA-DLA regimes, wakefield trapping can preaccelerate electrons before they exchange energy directly with the laser field \cite{Shaw2017, King2021}. In near-critical-density or relativistically transparent plasmas, ion motion and self-generated electromagnetic fields have been shown to affect the injection trajectories of electrons entering the laser-driven channel \cite{Jiang2018}. Other studies have considered electron supply from preformed channel walls, engineered guiding structures, weakly ionized high-Z plasmas, or externally modified channel fields \cite{Ma2018, Cohen2024, Valenta2024}. These studies demonstrate that DLA injection can be strongly influenced by the surrounding plasma structure and self-consistent fields. However, they also indicate that the injection picture remains fragmented. Existing explanations often rely on particular plasma configurations or parameter regimes, and they do not yet provide a general physical picture for how electrons are continuously supplied to a relativistic laser-driven plasma channel during propagation.

This limitation is particularly important for long-pulse relativistic laser propagation in plasma, \re{where the pulse front acts not merely the leading edge of the envelope but as a dynamically evolving interaction region.} 
% Previous studies have shown that the pulse front can undergo etching, steepening, localized energy depletion, and density compression, thereby strongly affecting laser propagation and energy dissipation \cite{Decker1996, Streeter2018}. However, these front-evolution effects have mainly been discussed in the context of pulse dynamics, rather than as part of the electron injection problem. As a result, a basic connection remains missing between the evolving laser front, the local restructuring of plasma-channel fields, and the sustained supply of electrons into the DLA phase. Establishing this connection is essential for explaining how DLA injection is maintained during laser propagation and why the electrons that eventually undergo DLA originate from organized, rather than randomly distributed, initial positions.
In this article, we propose a wavefront-guided electron injection mechanism for direct laser acceleration in plasmas. 
\re{We first characterize the propagation, etching, and local modulation of the relativistic pulse front during channel formation and analyze its propagation velocity in plasma. These front dynamics periodically reorganize the electron density and quasistatic fields near the channel entrance, producing spatially discrete regions favorable for electron injection.}
% \re{We argue that, as a relativistic laser pulse enters the plasma and drives the formation of a plasma channel, the propagation, etching, and local modulation of the pulse front periodically restructure the electron density and quasistatic-field distributions near the pulse front}. This process gives rise to spatially discrete injection regions in the vicinity of the channel entrance. 
\re{We then resolve the microscopic injection process, showing that electrons originating from favorable locations can acquire sufficient transverse momentum to escape the density pile-up layer at the pulse front, migrate into the channel, and subsequently re-enter the intense laser field to undergo DLA.}
% \re{Only electrons that originate from favorable initial locations and acquire sufficient transverse momentum can leave the wavefront pile-up layer under the action of these localized field structures}, move toward the interior of the channel, and eventually return to the intense-laser-field region where they enter the DLA phase. 
% Based on this physical picture, we further analyze the front velocity of a relativistic laser pulse propagating in plasma and its influence on the injection period. Using relativistic Particle-In-Cell simulations, \re{we verify the periodic character of electron injection and reveal its correspondence with laser-front modulation and self-generated magnetic island structures}. 
\re{Relativistic particle-in-cell simulations further reveal the periodic character of this process and establish its connection to the pulse-front propagation velocity, local front modulation, and self-generated magnetic-island structures.}
% Our results show that wavefront-guided electron injection is 
% %not an incidental feature of a specific simulation condition, but 
% an important mechanism that continuously supplies electrons for DLA in relativistic laser-driven plasma channels.
\re{These results identify wavefront-guided injection as an important mechanism for continuously supplying electrons to DLA in relativistic laser-driven plasma channels.}

\section{PIC simulations}
The simulations were performed using the fully relativistic Particle-In-Cell (PIC) code EPOCH~\cite{arber2015contemporary}. In the two-dimensional PIC simulations, a near-critical-density target is irradiated by a linearly polarized laser pulse, with the transverse electric field polarized along the y direction. The main case employs a laser pulse with a peak intensity of $I_0=1.0 \times 10^{21}\mathrm{W/cm}^2$. For a laser wavelength of $\lambda_0=2\pi c/\omega_0=800 nm$, this corresponds to the normalized amplitude of $a_0\equiv eE_l/(m_ec\omega_0)= 21.5$ with  and $c$ the speed of light. 
The laser has a focal spot size of $w_0=10\mathrm\mu m$ and a pulse duration of $\tau_0=200\mathrm fs$, both defined in terms of the full width at half maximum of the \re{Gaussian} profile. 
\re{The laser pulse propagates along the $x$ direction and is linearly polarized along the $y$ direction, corresponding to the transverse electric field $E_y$.} 
Since the present study focuses primarily on the electron injection process, the simulation is carried out for 1 ps. As will be shown below, the mechanism proposed here governs the continuous supply of electrons for direct laser acceleration, and this simulation duration is sufficient for the present analysis.
\begin{figure*}
\centering
\includegraphics[width=0.8\textwidth]{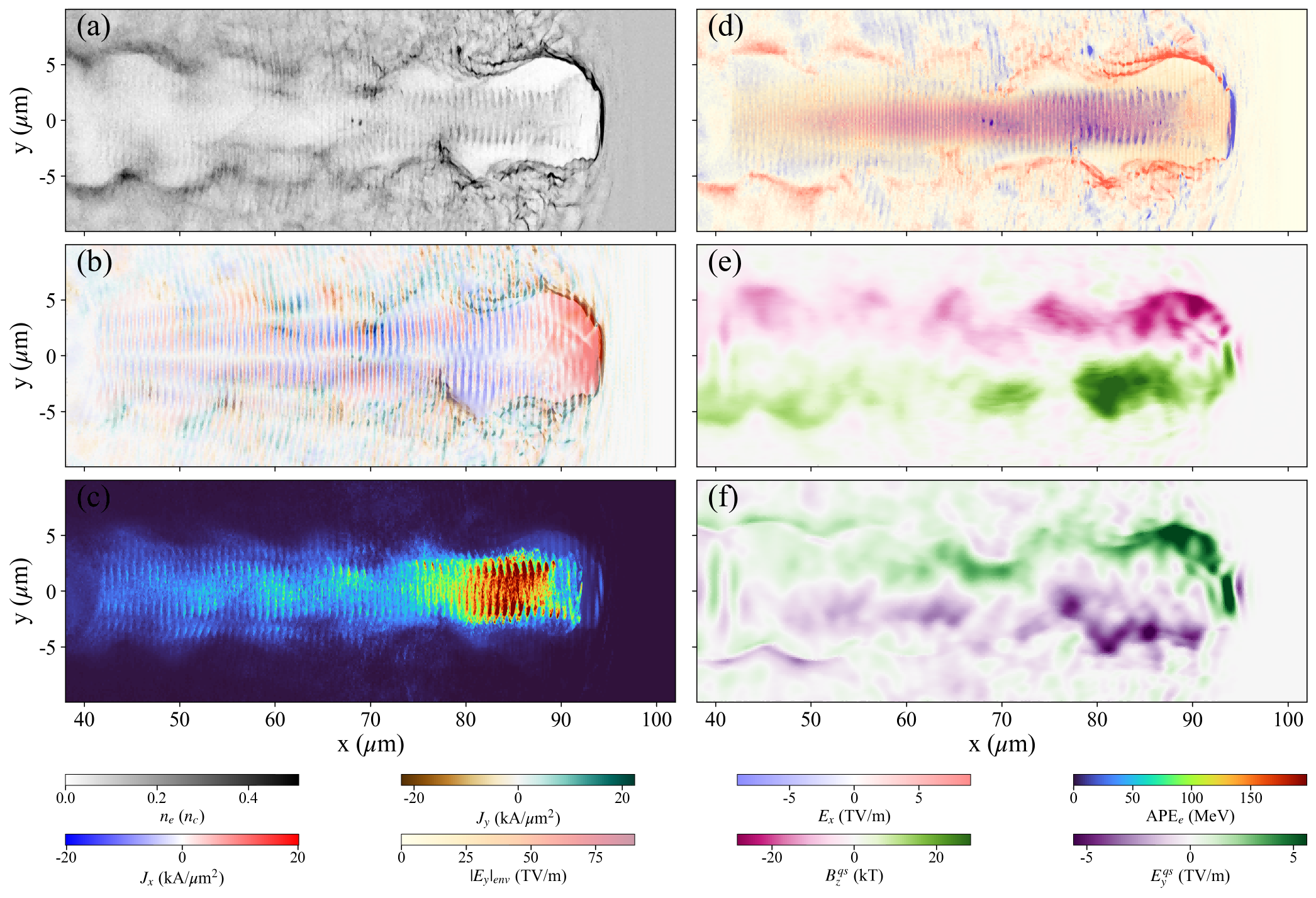}
\caption{2D PIC simulation results of $t=340\,fs$. The spatial distribution of (a) electron density $n_e$, (b) transverse current $J_y$ and longitudinal electric field $E_x$, (c) cell-averaged particle kinetic energy for electrons, (d) longitudinal current $J_x$ and electric field intensity of the laser envelope $|E_y|_{env}$, (e)quasistatic azimuthal magnetic field $B^{qs}_z$, and (f) quasistatic transverse electric field $E^{qs}_y$. }
\label{fig:field_340fs}
\end{figure*}

\begin{figure*}
\centering
\includegraphics[width=0.8\textwidth]{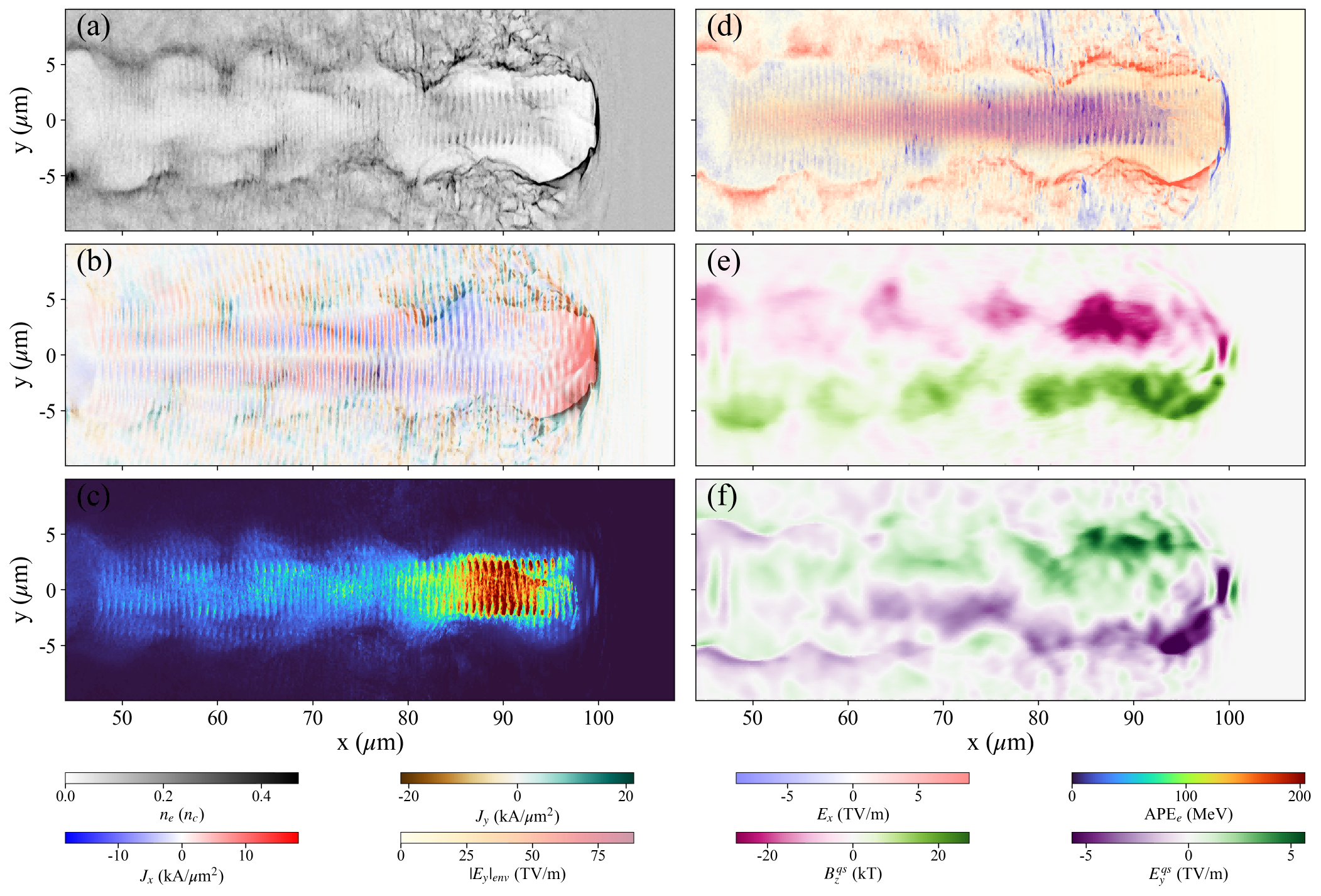}
\caption{2D PIC simulation results of $t=360\,fs$. The spatial distribution of (a) electron density $n_e$, (b) transverse current $J_y$ and longitudinal electric field $E_x$, (c) cell-averaged particle kinetic energy for electrons, (d) longitudinal current $J_x$ and electric field intensity of the laser envelope $|E_y|_{env}$, (e)quasistatic azimuthal magnetic field $B^{qs}_z$, and (f) quasistatic transverse electric field $E^{qs}_y$. }
\label{fig:field_360fs}
\end{figure*}

The simulation window has a size of $100 \mathrm{\mu m} \times 40 \mathrm{\mu m}$, discretized into $2500\times400$ cells. The corresponding cell sizes are $40\mathrm{nm}\times100\mathrm{nm}$, namely $\lambda_0/20\times\lambda_0/8$. 
The electron density of the helium plasma is $n_0=2\times{10}^{20}\mathrm{cm}^{-3}$, corresponding to approximately $0.116 \mathrm{n}_c$, and the Helium ion density is $n_i=1\times{10}^{20}\mathrm{cm}^{-3}$. \re{The plasma has a flat-top longitudinal density profile with a uniform electron density of $n_e = n_0$.} Additional simulations with different plasma densities were also performed for the parameter scan used to verify the scaling behavior, as discussed later. \re{Furthermore, the calculation of the time-averaged fields shown in Figs.~\ref{fig:laser_etching} and ~\ref{fig:birthmap} employ the simulation window of size $300 \mathrm{\mu m} \times 40 \mathrm{\mu m}$, without moving window.} Radiation-reaction module is not included in the simulations, because the laser intensity considered here is sufficiently low for quantum electrodynamics (QED) effects to be neglected. We note that radiation-reaction-induced electron injection can be an important injection pathway for DLA in the extreme-field regime \cite{Ji2014,gong2019radiation}. In the present work, however, we focus on electron injection driven by the conventional plasma response in the non-QED regime.

%These laboratory conditions are analogous to an FRB with a frequency $\omega\approx 1.4\,$GHz and an intensity $I_0\approx 10^{4}\,\mathrm{W/cm}^2$ that propagates through a plasma with a density $n_e\approx 1\times10^7\,\mathrm{cm}^{-3}$.

\section{Laser-driven plasma channel}

% \begin{widetext}
% \begin{table}[h]
% \end{table}
% \end{widetext}
When a relativistically intense laser pulse incident on an underdense plasma ($n_0 < n_c$), and when its pulse length $L=c\tau_L$ exceeds the plasma wavelength $\lambda_p=2\pi c/\omega_p$, the laser-plasma interaction is usually no longer characterized by the formation of a rear-closed bubble. Instead, it more closely corresponds to a long-pulse regime involving relativistic self-focusing, self-guiding, and ponderomotive channel formation. In this regime, the cycle-averaged ponderomotive force associated with the transverse laser-intensity gradient expels electrons radially from the high-intensity region, leading to electron-density depletion near the laser propagation axis. At the same time, the relativistic mass increase reduces the local effective plasma frequency and modifies the transverse refractive-index profile. Together with the ponderomotive expulsion of electrons, this effect promotes laser self-focusing and self-channeling\cite{Hafizi2000, Chen1998}. Fig.~\ref{fig:field_340fs}(a) and Fig.~\ref{fig:field_360fs}(a) show that the distribution of electron density at two representative simulation times. The electrons near the center are expelled, leading to the formation of a clearly defined plasma channel.

% \begin{figure*}
% \centering
% \includegraphics[width=0.9\textwidth]{res_den_Exyz.png}
% \caption{PIC simulation results. The spatial distribution of (a) electron density $n_e$, (b) ion density $n_i$, (c) longitudinal electric field $E_x$, (d) transverse electric field $E_y$, and (e) EM field $E_z$. Here, the three columns present the results at the time of $\omega_0t/2\pi = 150$, $300$, and $450$.}
% \label{fig:res_den_Exyz}
% \end{figure*}

% During channel formation, electron evacuation is not simply a process in which electrons are pushed away by the laser. Rather, it is determined by the competition between the outward ponderomotive force and the restoring force produced by charge separation. 
\re{The formation of plasma channel is determined by the competition between the outward ponderomotive force and the restoring force produced by charge separation.}
The transverse ponderomotive force associated with the laser envelope can be approximated as
\begin{equation}
\label{eq:ponderomotive_force}
\mathbf{F}_{p} = -m_e c^2 \nabla \gamma_{\perp},
\end{equation}
where
\begin{equation}
\label{eq:gamma_perp}
\gamma_{\perp} \simeq \sqrt{1 + \frac{a^2}{2}},
\end{equation}
and $a$ is the normalized vector potential. 
% This force pushes electrons from the strong-field region toward the channel boundary. After the electrons leave the axis, the much heavier ions remain approximately immobile on short time scales, forming a positive charge background and establishing a radial electrostatic restoring field directed toward the axis. The resulting channel structure is therefore determined by the laser intensity, focal spot size, plasma density, pulse duration, the laser power relative to the critical power for relativistic self-focusing, and by whether self-modulation, filamentation, or localized depletion occurs during propagation \cite{Esarey2009}.

In the near-axis region of an ideal, fully evacuated, axisymmetric ion channel, the radial electric field can be obtained from Gauss's law as
\begin{equation}
\label{eq:radial_electric_field}
E_r(r) \simeq \frac{e n_i}{2\epsilon_0}r,
\end{equation}
where $n_i$ is the ion density. For an electron, this field provides a restoring force directed toward the channel axis,
\begin{equation}
\label{eq:restoring_force}
F_r = -eE_r \simeq -\frac{m_e\omega_p^2}{2}r.
\end{equation}

The transverse motion of a near-axis electron can therefore be approximated as a relativistically corrected harmonic oscillation, with the betatron frequency given by
\begin{equation}
\label{eq:betatron_frequency}
\omega_{\beta} \simeq \frac{\omega_p}{\sqrt{2\gamma}}.
\end{equation}

% This near-axis linear focusing model provides a basic starting point for ion-channel betatron motion and theories of direct laser acceleration \cite{Arefiev2015}. 

Laser-driven plasma channels provide a natural environment for direct laser acceleration (DLA) of electrons\cite{Arefiev2015}, as has been demonstrated in simulations of relativistic laser channels and in self-channeling experiments \cite{Pukhov1999,Gahn1999,Mangles2005,Hussein2021}. In such a structure, the channel fields transversely confine electrons and drive betatron-like oscillations, while also allowing oscillating electrons to maintain spatial overlap with the laser field over an extended distance. 
The electron energy gain is therefore primarily determined by the work done by the laser field, namely
\begin{equation}
\label{eq:energy_gain_rate}
\frac{d\gamma}{dt}
= -\frac{e}{m_e c^2}\mathbf{v}\cdot\mathbf{E}.
\end{equation}

\begin{figure*}
\centering
\includegraphics[width=0.8\textwidth]{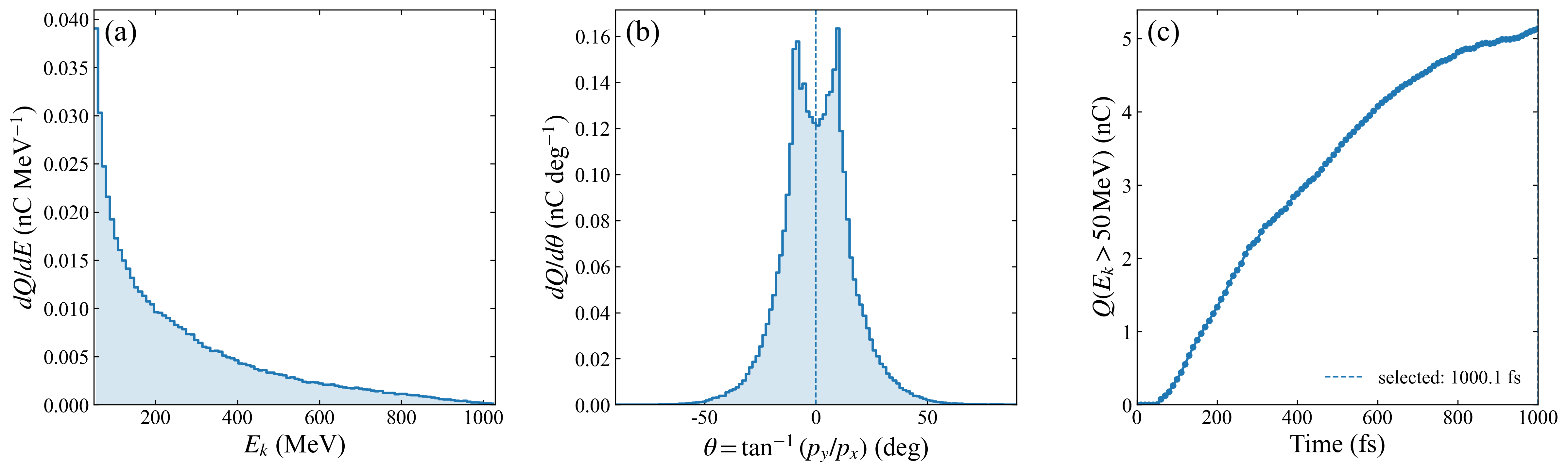}
\caption{2D PIC simulation results. The energy spectrum (a), angular distribution (b) of energetic electrons at 1 ps. (c) Charge evolution of the high-energy electrons ($E_k > 50 MeV$).}
\label{fig:output}
\end{figure*}

In particular, the transverse laser-field contribution,
\begin{equation}
\label{eq:transverse_laser_work}
-\frac{e}{m_e c^2}v_yE_y,
\end{equation}
usually provides the dominant contribution to the DLA energy gain\cite{Pukhov1999,Arefiev2012,Robinson2013}. When the channel-induced betatron motion maintains a favorable phase relation between the electron transverse velocity and the laser electric field, electrons can continuously extract net energy from the transverse laser field over multiple laser periods \cite{Pukhov1999,Gahn1999,Arefiev2012,Hussein2021}. Equivalently, the betatron oscillation modulates the dephasing of electrons with respect to the laser wave, allowing the cycle-averaged laser work to remain positive. This condition is commonly expressed as a phase-synchronization relation between the betatron frequency and the Doppler-shifted laser frequency,
\begin{equation}
\label{eq:dla_resonance_condition}
N\omega_{\beta}
\simeq
\left\langle
\omega_0\left(1 - \frac{v_x}{v_{\rm ph}}\right)
\right\rangle,
\end{equation}
where $N$ is an integer, $\omega_{\beta}$ is the betatron oscillation frequency, $\omega_0$ is the laser frequency, $v_x$ is the longitudinal electron velocity, and $v_{\rm ph}$ is the laser phase velocity in the plasma channel\cite{Pukhov1999,Arefiev2012,Hussein2021}. Therefore, in our simulations, once a long laser pulse self-consistently forms a plasma channel in the uniform plasma, the channel structure simultaneously provides transverse confinement, betatron oscillations, laser-electron spatial overlap, and dephasing modulation. The concurrent fulfillment of these conditions enables a large number of electrons to naturally enter the DLA process and be accelerated to high energies. Fig.~\ref{fig:field_340fs}(a) and ~\ref{fig:field_340fs}(c) show the formation of DLA electron beams in the channel, with the main portion of the bunch already reaching an average energy above $100\,\mathrm{MeV}$. \re{Fig.~\ref{fig:output} shows the parameters of the high-energy ($E_k > 50 MeV$) electrons obtained in the simulation. The calculation of charge is based on the assumption that the 2D PIC simulation has an effective thickness of 1 $\mu m$ in the unanalyzed $z$ direction.}

In realistic laser-driven plasma channels, the quasistatic fields inside the channel do not follow an ideal linear profile because of the nonlinear effects associated with laser propagation\cite{Meir2025, Yi2026}. Fig.~\ref{fig:field_340fs}(e), ~\ref{fig:field_340fs}(f), ~\ref{fig:field_360fs}(e) and ~\ref{fig:field_360fs}(f) show the distributions of the quasistatic azimuthal magnetic field and quasistatic transverse electric field in the plasma channel at two representative simulation times. The quasistatic fields are extracted from the PIC field data using a smoothed low-pass filter. These field maps reveal a pronounced spatial nonuniformity of the channel fields. Notably, although the field directions are antisymmetric with respect to y=0, the field amplitudes are not symmetrically distributed about the channel axis. In particular, within a region approximately $5\,\mathrm{\mu m}$ to $15\,\mathrm{\mu m}$ behind the channel front, a localized enhancement of the channel fields appears on one side of the axis. This localized enhancement is periodically redistributed as the laser propagates. At $t=340\,\mathrm{fs}$, corresponding to Fig.~\ref{fig:field_340fs}, the enhanced field region is located at $y<0$, whereas at $t=360\,\mathrm{fs}$, corresponding to Fig.~\ref{fig:field_360fs}, it appears at $y>0$. 
% As pointed out by Arefiev et al., quasistatic transverse electric fields, longitudinal electric fields, injection dynamics, and quasistatic magnetic fields can all modify the phase evolution and energy gain of electrons interacting with the laser field \cite{Arefiev2015}. 
Electron dynamics in realistic plasma channels are therefore generally more complex than those described by the idealized linear ion-channel model.

In addition to the channel-guided DLA process discussed above, Fig.~\ref{fig:field_340fs} and ~\ref{fig:field_360fs} also show a pronounced electron-density pile-up layer near the front of the laser-driven channel. This structure should be understood as a characteristic feature of the nonlinear propagation of an ultraintense laser pulse in an underdense plasma. For a relativistically intense laser pulse, the pulse front is strongly coupled to the plasma through the ponderomotive force. It excites a large-amplitude plasma response and locally transfers laser energy to plasma waves and electron motion. This localized energy transfer leads to pulse-front etching, in which the leading edge of the laser pulse is preferentially depleted and recedes backward relative to the main body of the pulse during propagation \cite{Decker1996,Streeter2018}.
\re{The formation of this density perturbation is consistent with the classical physical picture of ultraintense laser propagation in underdense plasmas \cite{Decker1996,Streeter2018}. The density pile-up observed at the channel front in Fig.~\ref{fig:field_340fs} and ~\ref{fig:field_360fs} should therefore be interpreted as a self-consistent front-layer structure of the laser-driven plasma channel.}
\re{As will be shown below, the observed density pile-up layer therefore establishes a physically grounded link between nonlinear pulse front evolution and the subsequent selection of high-energy DLA electrons.}

If pulse evolution is neglected, the propagation velocity of the laser envelope in an underdense plasma can be obtained from the linear dispersion relation as
\begin{equation}
\label{eq:linear_group_velocity}
v_g
= c\sqrt{1 - \frac{\omega_p^2}{\omega_0^2}}
= c\sqrt{1 - \frac{n_e}{n_c}}.
\end{equation}
where $\omega_p$ and $\omega_0$ are the plasma frequency and the laser frequency, respectively, and $n_c$ is the critical density corresponding to $\omega_0$. \re{
The classical velocity of pulse front etching in the group velocity frame of the laser can be estimated as\cite{Decker1996, Streeter2018}}
\begin{equation}
\label{eq:etching_velocity_classical}
v_{\rm etch}
\approx c\frac{\omega_p^2}{\omega_0^2}
= c\frac{n_e}{n_c}.
\end{equation}

Thus, in the laboratory frame, the effective velocity of the front that sustains the strong-field drive is commonly approximated as
\begin{equation}
\label{eq:front_velocity_classical}
v_{\rm front} \approx v_g - v_{\rm etch}.
\end{equation}

In the underdense limit $\omega_p \ll \omega_0$, expanding both $v_g$ and $v_{\rm etch}$ gives
\begin{equation}
\label{eq:front_velocity_underdense}
v_{\rm front}
\approx c\left(1 - \frac{3}{2}\frac{\omega_p^2}{\omega_0^2}\right).
\end{equation}

For a plasma channel driven by a strongly relativistic laser field, especially in near-critical-density plasmas, the above estimate needs to be modified to account for relativistic effects. The laser field increases the effective relativistic mass of electrons, thereby reducing the effective plasma frequency \cite{Weng2012}. 
As a result, the transparency condition is modified from $n_e < n_c$ to approximately $n_e < \gamma_{\rm eff}n_c$. We therefore introduce
\begin{equation}
\label{eq:epsilon_definition}
\epsilon \equiv \frac{n_e}{\gamma_{\rm eff}n_c}, \qquad \epsilon < 1,
\end{equation}
and write the local dispersion relation as
\begin{equation}
\label{eq:relativistic_dispersion}
\omega^2 = \frac{\omega_p^2}{\gamma_{\rm eff}} + c^2k^2.
\end{equation}

The corresponding relativistically corrected group velocity is then determined.
\begin{equation}
\label{eq:relativistic_group_velocity}
v_g
= c\sqrt{1 - \frac{\omega_p^2}{\gamma_{\rm eff}\omega_0^2}}
= c\sqrt{1 - \epsilon}.
\end{equation}

If the pulse-front etching scaling of Decker et al. is further modified by the relativistic transparency correction, an empirical estimate applicable to near-critical-density plasmas can be written as
\begin{equation}
\label{eq:etching_velocity_relativistic}
v_{\rm etch}
\approx c\frac{n_e}{\gamma_{\rm eff}n_c}
= c\epsilon.
\end{equation}

The effective front velocity can then be estimated as
\begin{equation}
\label{eq:front_velocity_ncd}
\begin{aligned}
v_{\rm front}
&\approx
c\sqrt{1 - \frac{n_e}{\gamma_{\rm eff}n_c}}
- c\frac{n_e}{\gamma_{\rm eff}n_c} \\
&= c\sqrt{1 - \epsilon} - c\epsilon.
\end{aligned}
\end{equation}

In the relativistic transparent limit $\epsilon \ll 1$, this expression becomes
\begin{equation}
\label{eq:front_velocity_ncd_limit}
\begin{aligned}
v_{{\rm front},\, {\rm tr}}
&\approx c\left(1 - \frac{3}{2}\epsilon\right) \\
&= c\left(1 - \frac{3}{2}\frac{n_e}{\gamma_{\rm eff}n_c}\right).
\end{aligned}
\end{equation}

\begin{figure}
\centering
\includegraphics[width=0.36\textwidth]{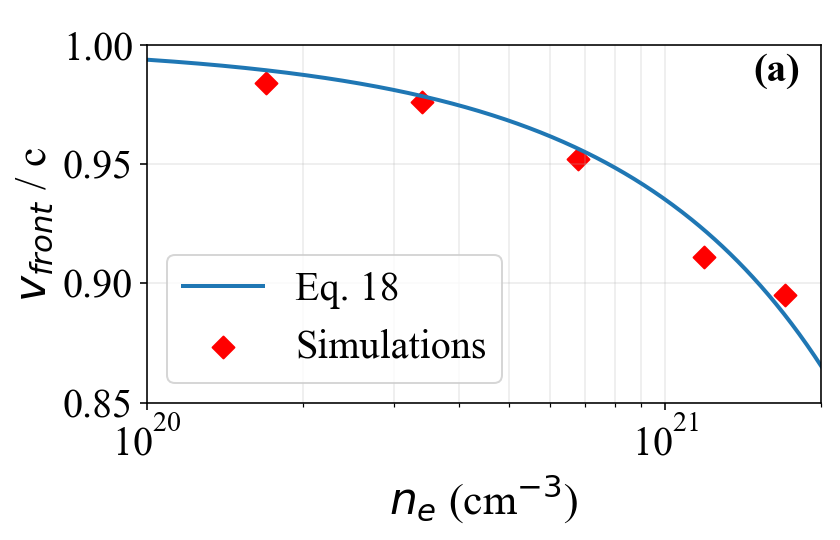}
\centering
\includegraphics[width=0.36\textwidth]{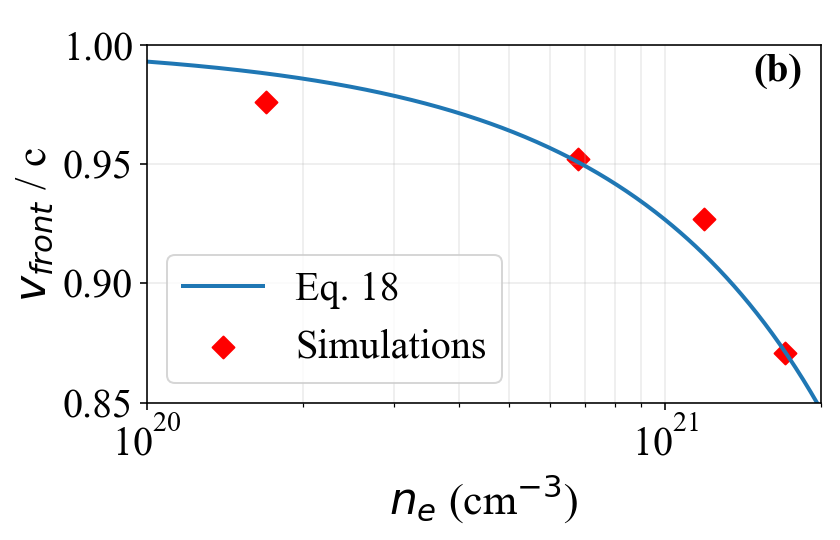}
\caption{The comparison between the effective front velocity calculated using Eq.~\ref{eq:front_velocity_ncd_limit} and the front velocity extracted from the PIC simulation in the range of near-critical density. (a) $a_0=9.6$; (b)$a_0=21.5$}
\label{fig:front_velocity}
\end{figure}

For a linearly polarized laser, the effective relativistic factor can be approximated as
$\gamma_{\rm eff} \approx \sqrt{1 + {a^2}/{2}}$. \re{We compare the front velocity estimated from the Eq.~\ref{eq:front_velocity_ncd_limit} with that extracted from the PIC simulations, as shown in Fig.\ref{fig:front_velocity}. This agreement indicates that the reduced propagation velocity of the channel front can be consistently described by the combined effects of pulse-front etching and relativistic transparency. }

In this etching-guided plasma channel, a large number of DLA electrons are injected into the channel. In the following section, we discuss the dynamics of electron injection based on the properties of the plasma channel.

\section{mechanism of electron injection}

\begin{figure*}
\centering
\includegraphics[width=0.8\textwidth]{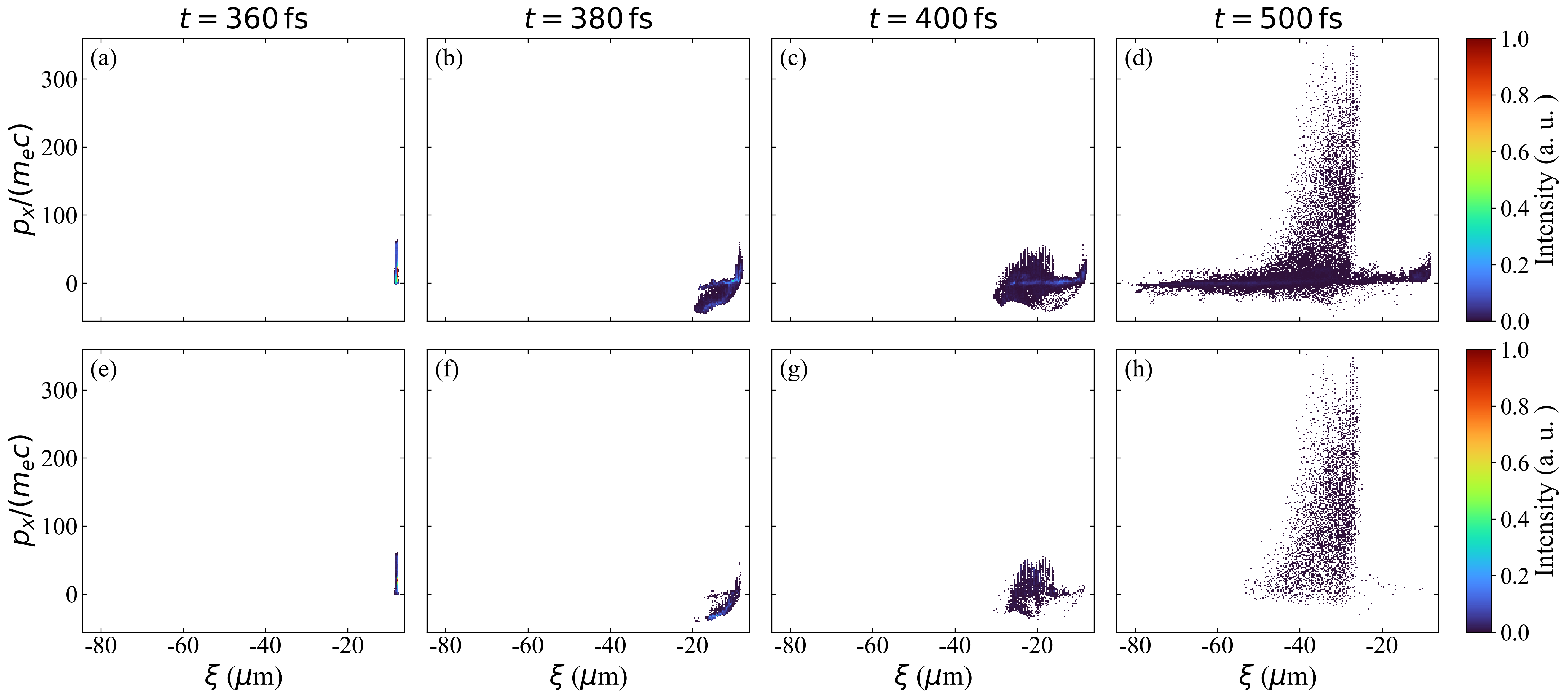}
\caption{\re{Temporal evolution of the longitudinal phase space of the electrons within the pile-up layer in Fig.~\ref{fig:field_360fs}. The top row shows all electrons in the pile-up layer, while the bottom row shows the injected electrons. From left to right, the panels correspond to $t= 360fs$ (a, e), $t= 380fs$ (b, f), $t= 400fs$ (c, g), and $t= 500fs$ (d, h), respectively.}}
\label{fig:x_phase_space_360fs}
\end{figure*}

\begin{figure*}
\centering
\includegraphics[width=1\textwidth]{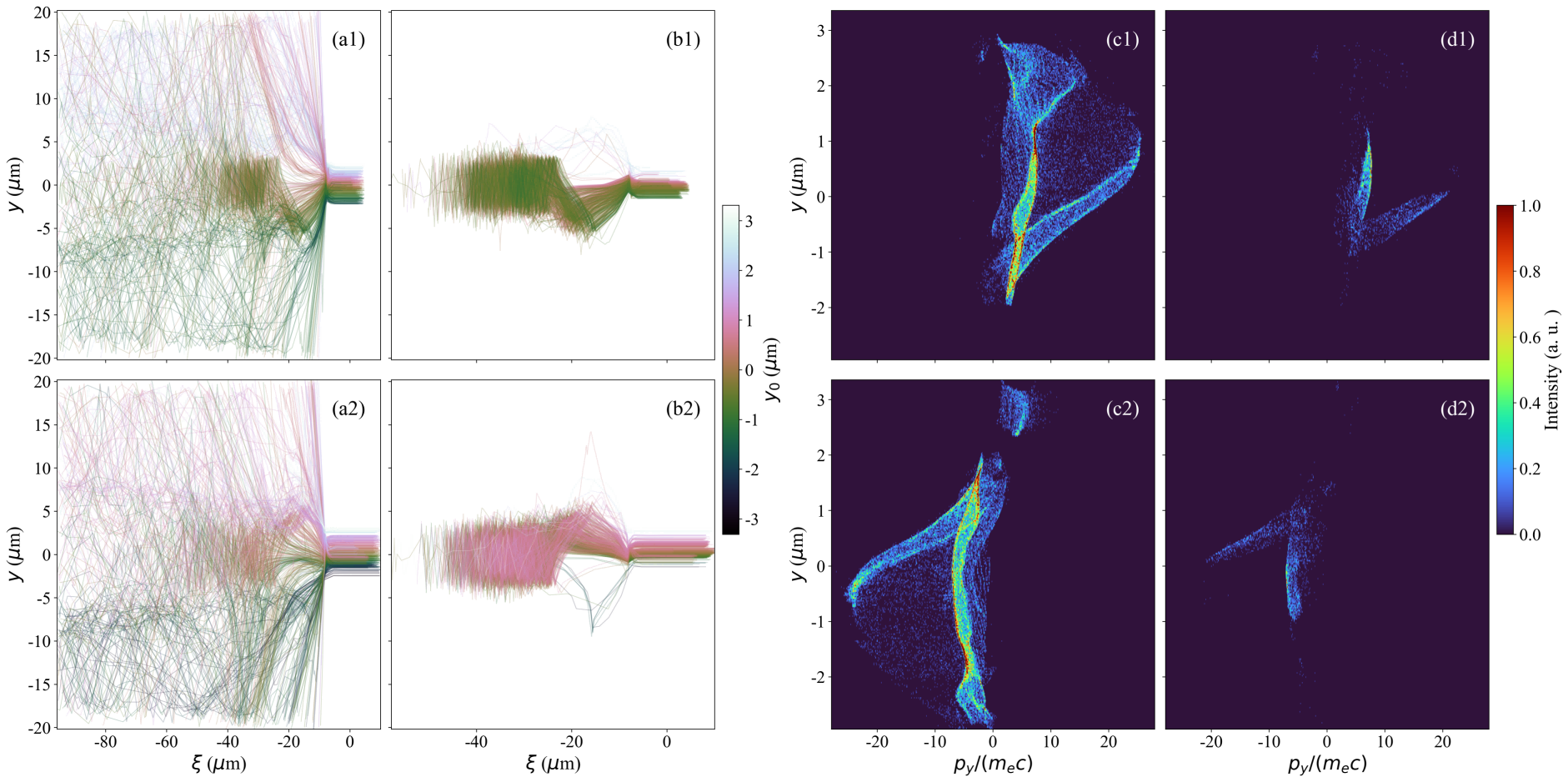}
\caption{Particle tracking of electrons in the pile-up layer at a specified time: (a) Trajectory of randomly sampled electrons among all electrons in the comoving frame; (b) Trajectory of randomly sampled electrons among the high-energy injected electrons (kinetic energy exceeding 200 MeV at 1 ps) in the comoving frame; (c) The transverse phase space distribution of all electrons in the density pile-up layer; (d) The transverse phase space distribution of high-energy injected electrons (kinetic energy exceeding 200 MeV at 1 ps) in the density pile-up layer. The first row (a1)(b1)(c1)(d1) corresponds to the simulation time of 340 fs, and the second row  (a2)(b2)(c2)(d2) corresponds to the simulation time of 360 fs.}
\label{fig:track_pile-up}
\end{figure*}

In the injection mechanism considered here, the electrons that are eventually injected first enter the density pile-up layer at the laser front and only subsequently move into the plasma channel. This density pile-up layer shown in Fig.~\ref{fig:field_340fs} is characterized by a strongly enhanced electron density localized at the front, together with a negative longitudinal current density. We select this region from the PIC simulations and track the trajectories of electrons located inside the density pile-up layer before and after their interaction with this front-layer structure. 
\re{Fig.~\ref{fig:x_phase_space_360fs} shows the temporal evolution of the longitudinal phase space distribution of electrons within the pile-up layer in Fig.~\ref{fig:field_360fs}. The results of the $(\xi, p_x)$ phase space in the corresponding pile-up layer in Fig.~\ref{fig:field_340fs} is highly similar to the result shown in Fig.~\ref{fig:x_phase_space_360fs}.}
Fig.~\ref{fig:track_pile-up}(a) and ~\ref{fig:track_pile-up}(b) show representative electron trajectories obtained from the PIC simulations. These trajectories indicate that the electrons injected into DLA from the density pile-up layer exhibit a highly coherent dynamical behavior. Fig.~\ref{fig:track_pile-up}(c) and ~\ref{fig:track_pile-up}(d) further compare the transverse phase space of all electrons in the density pile-up layer with that of the subset of electrons that are subsequently injected and accelerated to high energies.

The electron dynamics in the electromagnetic field is described by the following equations:

\begin{equation}
\label{eq:lorentz_p}
\frac{d\mathbf{p}}{dt} = - |e|\mathbf{E} - \frac{|e|}{\gamma m_{e}c}\left[ \mathbf{p} \times \mathbf{B} \right]
\end{equation}

\begin{equation}
\label{eq:velocity}
\frac{d\mathbf{r}}{dt} = \frac{c}{\gamma}\frac{\mathbf{p}}{m_{e}c}
\end{equation}

\begin{equation}
\label{eq:energy_gain}
\frac{d\gamma}{dt} = - \frac{e}{m_{e}c^{2}}\mathbf{E} \cdot \mathbf{v}
\end{equation}

Within the density pile-up layer at the \re{pulse} front, the electromagnetic fields are highly nonlinear and spatially localized. Electrons in this layer experience a strong transverse force associated with the laser-front field structure. For the two-dimensional geometry considered here, the dominant transverse equation of motion can be written as

\begin{equation}
\label{eq:transverse_motion}
\frac{dp_{y}}{dt} = - e\left( E_{y} - v_{x}B_{z} \right)
\end{equation}

We decompose the transverse force into two contributions,

\begin{equation}
\label{eq:force_decomposition}
F_{y} = F_{y}^{ch} + F_{y}^{L}.
\end{equation}

where \(F_{y}^{ch}\) denotes the quasistatic transverse force associated with the plasma channel, and \(F_{y}^{L}\) denotes the high-frequency oscillatory force associated with the intense laser field encountered subsequently. During the stage in which electrons are guided by the channel field and gradually approach the conditions for DLA resonance, the direct laser acceleration gain mechanism has not yet become dominant, because significant phase locking with the laser field has not been established.

The injection process observed in the simulations suggests that the electrons are not directly trapped into the DLA orbit from the \re{pulse} front. Instead, injection proceeds through a transverse transport process consisting of four stages: interaction \re{within the pile-up layer}, transverse drift, redirection by the localized quasistatic-field region, and return to the near-axis region.

\re{During their passage through the pile-up layer at the laser-pulse front, whose longitudinal width is approximately one laser wavelength}, the electrons experience a short but strong transverse impulse. The transverse momentum acquired after this stage can be written as

\begin{equation}
\label{eq:pin_def}
p_{in}(\xi_{0}) = \int_{\Delta t_{sp}}^{}F_{y}^{sp}(\xi_{0},t)\,dt,
\end{equation}

where

\begin{equation}
\label{eq:fsp_def}
F_{y}^{sp} = - e(E_{y} - v_{x}B_{z})_{sp}.
\end{equation}

% Because the pile-up layer is phase-locked to the laser front, this impulse depends on the initial longitudinal coordinate \(\xi_{0}\). 
\re{Because the pile-up layer remains localized at the laser-pulse front, the laser-field phase encountered by an electron, and hence the transverse impulse it receives, depends on its initial comoving coordinate $\xi_0$.}
Thus,

\begin{equation}
\label{eq:pin_xi}
p_{in} = p_{in}(\xi_{0}).
\end{equation}

In addition to imparting transverse momentum, the same impulse also produces a small transverse displacement during the pile-up interaction,

\begin{equation}
\label{eq:delta_y_sp}
\Delta y_{sp}(\xi_{0}) = \int_{\Delta t_{sp}}^{}\frac{p_{y}(t)}{\gamma m_{e}}\,dt.
\end{equation}

Under a short-impulse approximation, this displacement scales with the acquired transverse momentum,

\begin{equation}
\label{eq:delta_y_sp_scaling}
\Delta y_{sp}(\xi_{0}) \simeq \chi_{sp}p_{in}(\xi_{0}),
\end{equation}

where \(\chi_{sp}\) is an effective coefficient determined by the pile-up interaction time and the characteristic electron energy during this stage. Therefore, after the pile-up interaction, the electron enters the subsequent channel-transport stage with

\begin{equation}
\label{eq:yin}
y_{in} = y_{0} + \Delta y_{sp}(\xi_{0}),
\end{equation}

\begin{equation}
\label{eq:pyin}
p_{y,in} = p_{in}(\xi_{0}).
\end{equation}

% This step demonstrates that both the injection position and the injection momentum are modulated by the same laser-front phase $\xi_{0}$.
\re{This analysis demonstrates that the initial position and the initial transverse momentum exhibit a common dependence on $\xi_0$, reflecting their modulation by the local laser-field phase within the pulse-front region. }By comparing the phase-space distributions of electrons in the density pile-up layer at two different times, shown in Fig.~\ref{fig:track_pile-up}(c1) and ~\ref{fig:track_pile-up}(\re{c2}), we can observe clear differences in the electron momentum distribution. 

\begin{figure}
\centering
\includegraphics[width=0.48\textwidth]{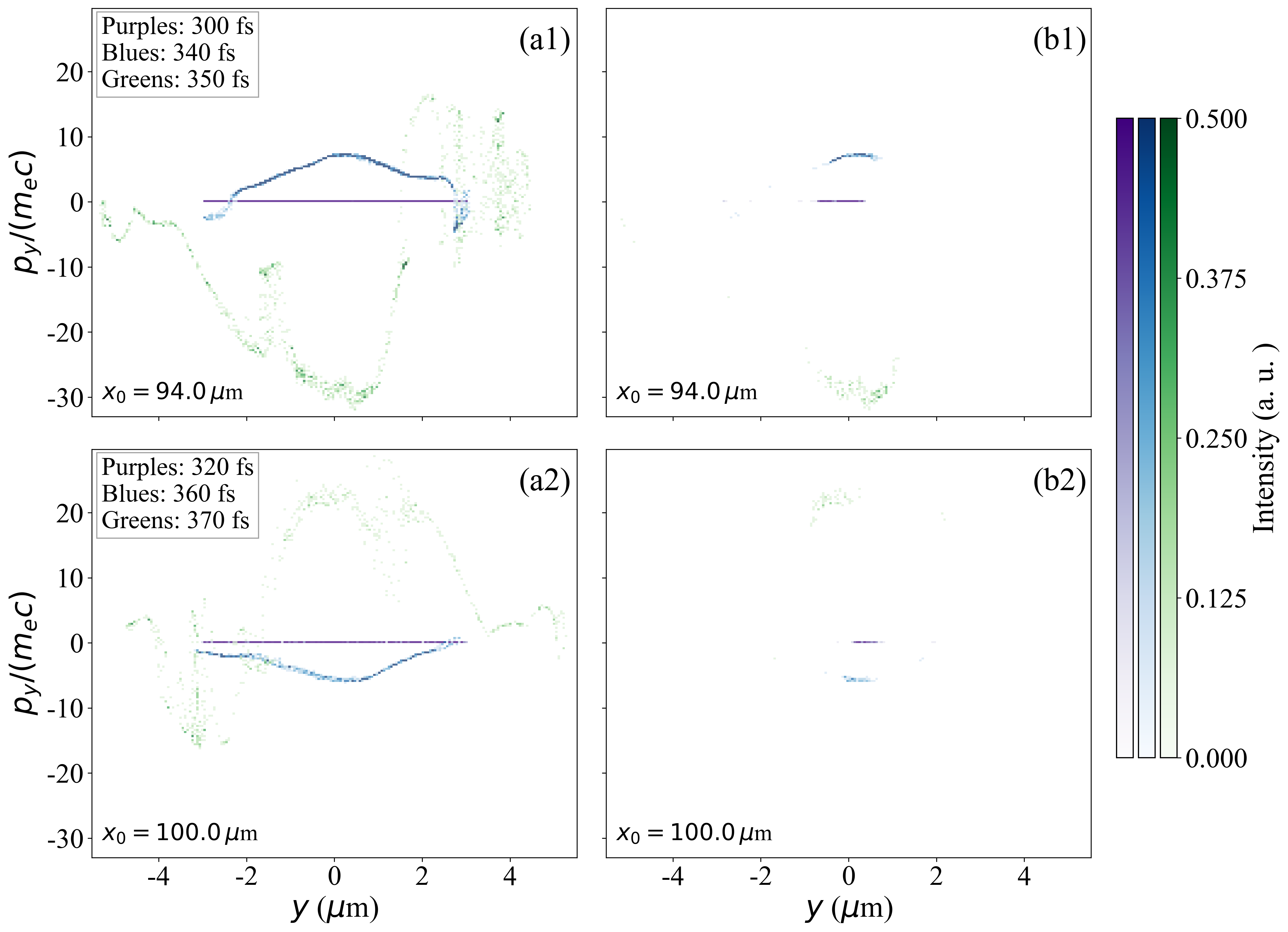}
% \centering
% \includegraphics[width=0.48\textwidth]{fig5b_no.png}
\caption{The transverse phase space distribution of electrons in x-slice of initial plasma, after the formation of the density pile-up layer (Stage 1): (a) all the electrons in the x-slice, (b) the injected electrons in the x-slice (final energy exceeds 100 MeV at 1 ps). First row (a1)(b1): $x=94\,\mu m$ slice ; second row (a2)(b2): $x=100\,\mu m$ slice.}
\label{fig:phase_space}
\end{figure}
To further clarify how different values of \(\xi_{0}\) affect \(y_{in}\) and \(p_{y,in}\) in the simulation, Fig.~\ref{fig:phase_space} shows the transverse phase-space distributions of electrons from two selected \(x\)-slices before and after they pass through the density pile-up layer, corresponding to different values of \(\xi_{0}\). In Fig.~\ref{fig:phase_space}, the purple distribution corresponds to the electron phase space before the laser arrives, the blue distribution corresponds to the phase space when the electrons are located inside the density pile-up layer, and the green distribution corresponds to the phase space when the electrons have just entered the channel field from the pile-up layer, namely at the initial stage of outward transverse drift.
In addition to the information shown in Fig.~\ref{fig:phase_space}, we also selected electrons from the same \(x\)-slices using different high-energy thresholds at \(t = 1\,ps\) based on trajectory tracking. The injectable phase-space region remains nearly unchanged under these different thresholds. This indicates that the injection process is governed by a well-defined injectable window, rather than simply by a high-energy injection window. In Fig.~\ref{fig:phase_space}(b), we show the phase space of electrons selected with an energy threshold of \(100\,MeV\) at the corresponding time.

\begin{figure}
\centering
\includegraphics[width=0.48\textwidth]{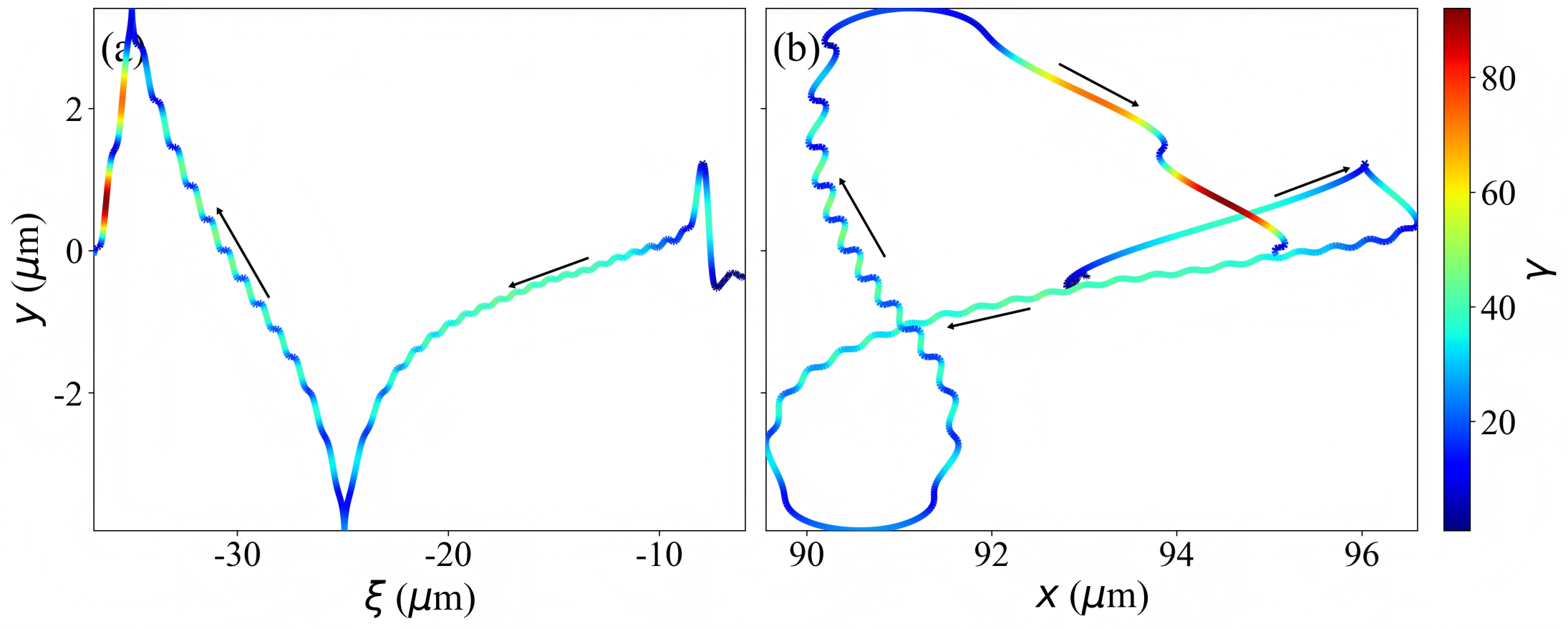}
\includegraphics[width=0.48\textwidth]{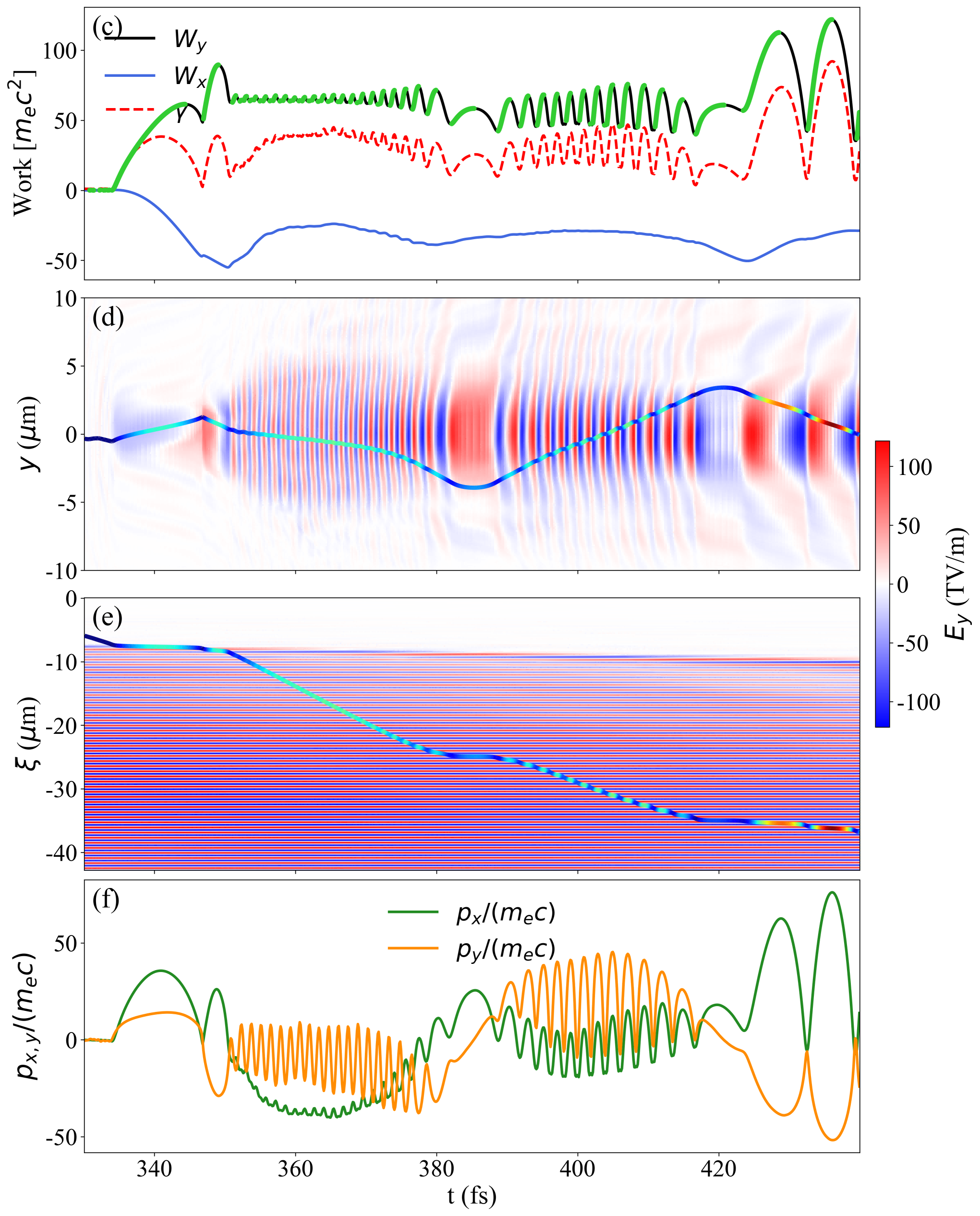}
\caption{Particle tracking of representative electron in the injection process from the PIC simulation.(a) The trajectory of comoving frame. (b) The trajectory of laboratory frame. (c) Work by transverse and longitudinal electric fields. (d) Transverse oscillations with respect to the x slice of $E_y$. The background color is $E_y$ at x slice where the electron is located at.(e)Electron trajectory in a window moving with the speed of light. The background color is $E_y$ at y = 0 and the vertical coordinate is $\xi = x - ct$. (f)The time evolution of transverse and longitudinal momentum. Note that the thick part of the curves in panel (c) (highlighted with green) indicates the part of the trajectory where the electron is gaining energy from $E_y$.
}
\label{fig:track_dashboard}
\end{figure}
By comparing the phase-space distributions at the three times, we find that the transverse momenta of all electrons oscillate with the laser phase over the interval \(\Delta t_{sp}\). This behavior corresponds to the momentum evolution of the representative electron in Fig.~\ref{fig:track_dashboard}(f) during its interaction with the first laser cycle. In Fig.~\ref{fig:phase_space}, the injected electrons in the two \(x\)-slices acquire transverse momenta with opposite signs after leaving the pile-up layer. In Fig.~\ref{fig:phase_space}(b1), the positions of the injected electrons at the blue and green times are shifted in the \(+ y\) direction relative to their initial positions \(y_{0}\). In Fig.~\ref{fig:phase_space}(b2), by contrast, the positions of the injected electrons at the blue and green times are shifted in the \(- y\) direction relative to \(y_{0}\).

Together, these observations show that the injected electron population satisfies a phase-dependent relation in both the injection momentum and the transverse displacement, namely \(p_{in}(\xi_{0})\) and \(\Delta y_{sp}(\xi_{0})\). Fig.~\ref{fig:phase_space} therefore demonstrates that the transverse phase-space conditions for injection are systematically controlled by \re{the local laser-field phase associated with the initial coordinate} \(\xi_{0}\).

The subsequent injection is governed by whether this entrance phase-space point allows the electron to reach the localized strong quasistatic-field region near the channel wall. This region is visible in the quasistatic field distribution (Fig.~\ref{fig:field_340fs} (e)and ~\ref{fig:field_340fs} (f)) and acts as a transverse redirection region. It does not act as a conservative reflecting wall in the usual mechanical sense. Rather, it provides a sufficiently large inward transverse impulse that reverses the electron transverse momentum and sends the electron back toward the axis.

Let

\begin{equation}
\label{eq:s_def}
s = \operatorname{sgn}[ p_{in}(\xi_{0})]
\end{equation}

denote the direction of the initial outward motion. Suppose the relevant strong quasistatic reflection region for this branch is centered at

\begin{equation}
\label{eq:reflection_center}
\left( \xi_{R},y_{R}^{(s)} \right)
\end{equation}
with transverse width \(\Delta_{R}\). During this drift stage, the electrons still oscillate strongly in the high-frequency laser field. However, before they enter the DLA resonance regime, the laser field does not provide them with a cycle-averaged net energy gain. Before reaching the reflection region, the electron approximately drifts transversely according to its \re{initial} momentum. In the comoving coordinate \(\xi = x - ct\), one may write

\begin{equation}
\label{eq:dy_dxi}
\frac{dy}{d\xi} = \frac{v_{y}}{v_{x} - c} \simeq \frac{p_{in}(\xi_{0})}{\gamma m_{e}(v_{x} - c)}.
\end{equation}

Thus, the transverse position of the electron when it reaches \(\xi = \xi_{R}\) is approximately

\begin{equation}
\label{eq:y_arrival}
y_{R}^{arr} \simeq y_{0} + \Delta y_{sp}(\xi_{0}) + \frac{p_{in}(\xi_{0})}{\gamma m_{e}(v_{x} - c)}(\xi_{R} - \xi_{0}).
\end{equation}

Effective injection requires that the electron trajectory intersect the localized strong-field reflection region,

\begin{equation}
\label{eq:intersection_condition}
\lvert y_{R}^{arr} - y_{R}^{(s)} \rvert < \Delta_{R}.
\end{equation}

This condition defines a moving injection ridge in the initial \(\left( \xi_{0},y_{0} \right)\) space. The center of this ridge \(y_{c}(\xi_{0})\) satisfies

\begin{equation}
\label{eq:ridge_center_condition}
y_{c}(\xi_{0}) + \Delta y_{sp}(\xi_{0}) + \frac{p_{in}(\xi_{0})}{\gamma m_{e}(v_{x} - c)}(\xi_{R} - \xi_{0}) = y_{R}^{(s)}.
\end{equation}

Therefore,

\begin{equation}
\label{eq:yc_def}
y_{c}(\xi_{0}) = y_{R}^{(s)} - \Delta y_{sp}(\xi_{0}) - \frac{p_{in}(\xi_{0})}{\gamma m_{e}(v_{x} - c)}(\xi_{R} - \xi_{0}).
\end{equation}

Using

\begin{equation}
\label{eq:delta_y_use}
\Delta y_{sp}(\xi_{0}) \simeq \chi_{sp}p_{in}(\xi_{0}),
\end{equation}

we obtain

\begin{equation}
\label{eq:yc_pin}
y_{c}(\xi_{0}) \simeq y_{R}^{(s)} - \left[ \chi_{sp}+\frac{\xi_{R} - \xi_{0}}{\gamma m_{e}(v_{x} - c)} \right] p_{in}(\xi_{0}).
\end{equation}

Over a local injection segment, the quantities \(y_{R}^{(s)}\), \(\xi_{R}\), \(\gamma\), and \(v_{x}\) vary slowly compared with the optical-scale oscillation of \(p_{in}(\xi_{0})\). The relation can therefore be simplified as

\begin{equation}
\label{eq:yc_simplified}
y_{c}(\xi_{0}) \simeq Y_{R} - \chi_{eff}p_{in}(\xi_{0}).
\end{equation}

This expression explains why the center of the injectable transverse slice oscillates consistently with the initial transverse momentum. Both are ultimately controlled by the laser phase at the pile-up layer.

The finite transverse width of the reflection region gives a finite injection band rather than a single line,

\begin{equation}
\label{eq:injection_band}
\lvert y_{0} - y_{c}(\xi_{0}) \rvert < \Delta y_{inj},
\end{equation}

where, to lowest order,

\begin{equation}
\label{eq:bandwidth_lowest}
\Delta y_{inj} \sim \Delta_{R}.
\end{equation}

Electrons whose initial transverse positions lie outside this band do not intersect the strong quasistatic reflection region. As a result, they do not accumulate enough inward transverse impulse to reverse \(p_{y}\), and they tend to escape transversely instead of returning to the near-axis region.

If the trajectory angle, momentum spread, and the longitudinal width of the reflection region are taken into account, the injection bandwidth can be written as

\begin{equation}
\label{eq:bandwidth_full}
\Delta y_{inj} \sim \Delta_{R} + \left| \frac{\partial y_{R}^{arr}}{\partial p_{in}} \right| \Delta p_{in} + \left| \frac{dy}{d\xi} \right| \Delta\xi_{R}.
\end{equation}

This expression indicates that the injection bandwidth is jointly determined by the finite spatial extent of the reflection region and the phase-space spread at the entrance.

We can now write the injectable window of the initial electron position as a compact main criterion,

\begin{equation}
\label{eq:main_injection_criterion}
\left| y_{0} + \Delta y_{sp}(\xi_{0}) + \frac{p_{in}(\xi_{0})}{\gamma m_{e}(v_{x} - c)}(\xi_{R} - \xi_{0}) - y_{R}^{(s)} \right| < \Delta_{R}
\end{equation}

This criterion corresponds to the injectable window at the initial time, shown by the purple distributions in Fig. ~\ref{fig:phase_space}(b) and ~\ref{fig:phase_space}(d).

The reflection condition can also be expressed in terms of the transverse impulse supplied by the quasistatic channel field. The channel contribution to the transverse momentum is

\begin{equation}
\label{eq:channel_impulse}
\Delta p_{y}^{ch} = \int_{R}^{}F_{y}^{ch}\,dt = - e\int_{R}^{}(E_{y}^{qs} - v_{x}B_{z}^{qs})\,dt.
\end{equation}

For an electron initially moving in the \(s\) direction, redirection requires

\begin{equation}
\label{eq:redirection_condition}
- s\Delta p_{y}^{ch} \gtrsim \lvert p_{in} \rvert .
\end{equation}

If the electron misses the localized reflection region, \(\lvert \Delta p_{y}^{ch} \rvert\) is too small to reverse the transverse momentum, and the electron continues to drift outward. If it intersects the reflection region, the accumulated inward impulse reverses \(p_{y}\), allowing the electron to return toward the axis. Electrons located farther outside the narrow injection band, or deviating from it, are therefore more likely to escape. 
% This is not because of transverse potential-energy conservation, but because their trajectories do not pass sufficiently effectively through the strong reflection region and thus fail to accumulate enough reverse transverse impulse. 
In addition to the electron-dynamical analysis based on the localized strong quasistatic-field region, the presence of the strong reflection region is also manifested by the concentrated occurrence of trajectory turning points within a finite interval around \(\xi = - 20\ \mu m\) in Fig.~\ref{fig:track_pile-up}.

To clearly illustrate the complete injection process, from the electron entering the pile-up layer to its return to the near-axis region, we perform detailed trajectory tracking for a representative electron selected from the injected population. The trajectory of this representative electron throughout the injection process is shown in Fig.~\ref{fig:track_dashboard}(a) and ~\ref{fig:track_dashboard}(b). The work done on the electron by the transverse and longitudinal electric fields is shown in Fig.~\ref{fig:track_dashboard}(c). In Fig.~\ref{fig:track_dashboard}(d) and ~\ref{fig:track_dashboard}(e), we show the relative position of the representative electron with respect to the laser field. The detailed momentum evolution is presented in Fig.~\ref{fig:track_dashboard}(f). The trajectory-tracking results further confirm the four-stage injection process described above.

This picture also explains the interruption of injection at certain laser phases. Since \(p_{in}(\xi_{0})\) is phase-dependent, it can become very small near the zero-crossing phase of the pile-up impulse. In that case, electrons cannot drift far enough to reach the localized reflection region within the available channel length and interaction time. A minimum transverse momentum is therefore required,

\begin{equation}
\label{eq:pmin}
\lvert p_{in} \rvert \gtrsim p_{\min} \sim \gamma m_{e}\frac{d_{R}}{\tau_{R}},
\end{equation}

where \(d_{R} = s(y_{R}^{(s)} - y_{in})\) is the transverse distance from the entrance position to the reflection region and \(\tau_{R}\) is the available drift time. When $p_{in}(\xi_{0}) \approx 0$, the entire longitudinal slice fails to reach the reflection region, leading to an interruption of injection.
Thus, the injection process requires a finite transverse momentum acceptance,

% \begin{equation}
% \label{eq:pin_zero}

% \end{equation}

\begin{equation}
\label{eq:momentum_acceptance}
\lvert p_{in}(\xi_{0}) \rvert \gtrsim p_{\min}.
\end{equation}

% In this model, the channel wall should not be interpreted as a conservative potential wall. Instead, it is a localized quasistatic-field region that redirects electrons in transverse phase space. The injection window arises because only a limited set of initial transverse positions and phase-dependent momenta allows electrons to intersect this region, acquire sufficient inward impulse, and return to the near-axis region. After this return, the electrons can overlap with the laser field for a longer time and enter the subsequent DLA stage.

In this way, we obtain a complete description of the electron injection dynamics for an arbitrary \re{initial coordinate $\xi_{0}$ of the pulse front}. It should also be emphasized that the entire injection process relies on the localized strong quasistatic-field region, which corresponds to the magnetic-island structure discussed in previous studies\cite{Gong2021,Cai2025}. In fact, both the strength and the transverse position of this localized strong quasistatic-field region evolve with \(\xi_{0}\). For example, as shown in Fig.~\ref{fig:field_340fs} and ~\ref{fig:field_360fs}, the enhanced region appears on opposite sides of the \(y\)axis. Nevertheless, its position relative to the \re{pulse} front, denoted by \(\xi_{R}\), remains relatively well defined, which may be related to the plasma wavelength.
\begin{figure*}
\centering
\includegraphics[width=1\textwidth]{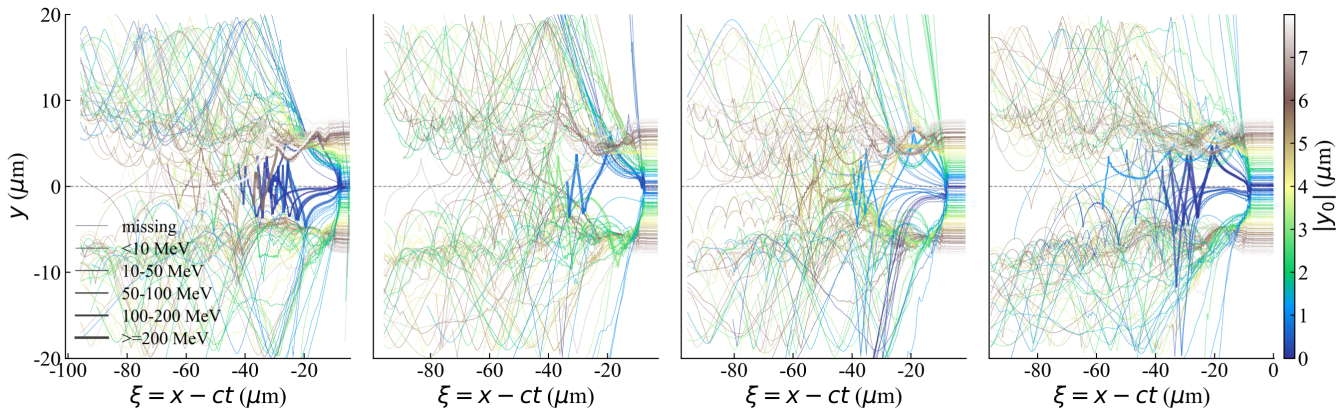}
\caption{The trajectory tracking diagram of the example electrons sampled uniformly from the x = 94.5 $\mu m$(a), 96.0 $\mu m$(b), 97.5 $\mu m$(c), 99.0 $\mu m$(d) slices in the comoving frame. Different line widths are used when drawing the electron tracks to distinguish the kinetic energy that electrons acquire at 1 ps.}
\label{fig:track_slice}
\end{figure*}
To illustrate how the variation of \(\xi_{0}\) affects electrons at different transverse positions before and after their interaction with the density pile-up layer, and in particular how the injection behavior changes with \(\xi_{0}\), Fig.~\ref{fig:track_slice} shows the trajectories of representative electrons that are initially uniformly distributed in the transverse direction on different x-slices, plotted in the comoving coordinate. Clear differences can be observed in the behavior of these uniformly sampled electrons for different values of \(\xi_{0}\).

In Fig.~\ref{fig:track_slice}(a), strong electron injection occurs. Electrons that acquire momentum in the -y direction after passing through the pile-up layer are reflected by the localized strong quasistatic-field region and are subsequently injected into the channel. In Fig.~\ref{fig:track_slice}(b) and ~\ref{fig:track_slice}(c), \(p_{in}(\xi_{0})\) reverses its sign, while both \(p_{in}(\xi_{0})\) and the localized strong quasistatic-field region gradually weaken. As a result, the injection process is interrupted. In Fig.~\ref{fig:track_slice}(d), the localized strong quasistatic-field region becomes enhanced again, but its enhanced region is flipped to the opposite side of the y=0 axis. Correspondingly, electron injection is also enhanced again. These results suggest that, as \(\xi_{0}\) varies periodically, the wavefront-guided electron injection process undergoes a corresponding periodic modulation.

Thus, using direct geometrical considerations together with the electron-dynamical facts discussed above, we obtain the relation given above. The pile-up layer imparts a phase-dependent momentum and displacement to the electrons, and the electrons must subsequently drift into the localized strong reflection region. To reach this reflection region, the original initial position \(y_{0}\) must compensate for these phase-dependent displacements. As a result, the center of the injectable window forms an injection ridge that oscillates with \(p_{in}(\xi_{0})\).

\section{birth map of the injected electrons}
Next, we consider the electron injection behavior during the entire process of laser-driven plasma-channel formation. Electrons in the density pile-up layer are mainly affected by the laser field at the \re{pulse front} and therefore acquire a corresponding transverse momentum. In a plane electromagnetic wave, the transverse momentum of an electron scales as $p_y \sim eA_y$. When estimating the electron momentum in the density pile-up layer, we therefore use the plane-wave approximation. The vector potential of the linearly polarized laser is written as
\begin{equation}
\label{eq:ay_plane_wave}
a_y = a_0\cos(\psi+\phi_0),
\end{equation}
where
\begin{equation}
\label{eq:laser_phase_xi}
\psi = \omega_0 t-k_0 x.
\end{equation}
Here, $\omega_0$ is the central laser frequency, and $k_0$ is the magnitude of the central laser wave vector. When $\phi_0=0$, the laser phase is simply $\psi=\omega_0 t-k_0 x$. The laser pulse front propagates with velocity $v_{\rm front}$, whereas the wavefront of the laser field propagates with phase velocity $v_{\rm ph}$. The density pile-up layer is attached to the \re{pulse} front. Therefore, its longitudinal position can be approximated as $x=v_{\rm front}t$.

When the density pile-up layer is located at $x$, the transverse momentum acquired by electrons at the front can be estimated as
\begin{equation}
\label{eq:py_front_integral}
\begin{aligned}
p_{y,{\rm front}}(x)
&\sim \int e a_0 \cos(\psi)\,\delta\left(t-\frac{x}{v_{\rm front}}\right)\,dt  \\
&= e a_0 \cos(k_{\rm lon}x),
\end{aligned}
\end{equation}
where
\begin{equation}
\label{eq:klon_definition}
k_{\rm lon}=k_0\frac{v_{\rm ph}-v_{\rm front}}{v_{\rm front}}.
\end{equation}
This result indicates that the transverse momentum of electrons in the density pile-up layer has a longitudinal spatial periodicity. We define the corresponding momentum-modulation wavelength as
\begin{equation}
\label{eq:lambda_lon}
\lambda_{\rm lon}=\frac{2\pi}{k_{\rm lon}}
=\lambda_0\frac{v_{\rm front}}{v_{\rm ph}-v_{\rm front}}.
\end{equation}
Including the initial laser phase $\phi_0$, the transverse momentum can be written as
\begin{equation}
\label{eq:py_front_phase}
p_{y,{\rm front}}(x)
\sim m_e c e a_0 \cos(k_{\rm lon}x+\phi_0).
\end{equation}
Substituting this expression into the center position of the electron-injection acceptance region gives
\begin{equation}
\label{eq:yc_periodic}
y_c(x)\simeq y_R^{(s)}
-m_e c e a_0\chi_{\rm eff}\cos(k_{\rm lon}x+\phi_0),
\end{equation}
where $\chi_{\rm eff}$ represents the effective mapping coefficient from the transverse momentum imparted at the front to the transverse displacement required for injection.

If the electron density in the pile-up layer is approximated as $n_0$, the transverse current modulation can be written with the same spatial periodicity as
\begin{equation}
\label{eq:jy_modulation}
j_y\simeq e n_0\cos(k_{\rm lon}x+\phi_0).
\end{equation}
This transverse-current modulation can then induce a longitudinally periodic structure in the quasistatic magnetic field of the plasma channel, with a period corresponding to that of the current modulation \cite{Gong2021}. Therefore, in the electron-injection problem, the initial spatial distribution of injected electrons is expected to correlate with the periodic distribution of the quasistatic magnetic field.

Using
\begin{equation}
\label{eq:ampere_bz_jy}
\frac{\partial \bar{B}_{z,{\rm lon}}}{\partial x}
=-\mu_0 j_y,
\end{equation}
we obtain
\begin{equation}
\label{eq:bz_lon_periodic}
\bar{B}_{z,{\rm lon}}
\sim -\frac{\mu_0 |e| n_0}{k_{\rm lon}}
\sin(k_{\rm lon}x+\phi_0).
\end{equation}

\begin{figure}
\centering
\includegraphics[width=0.48\textwidth]{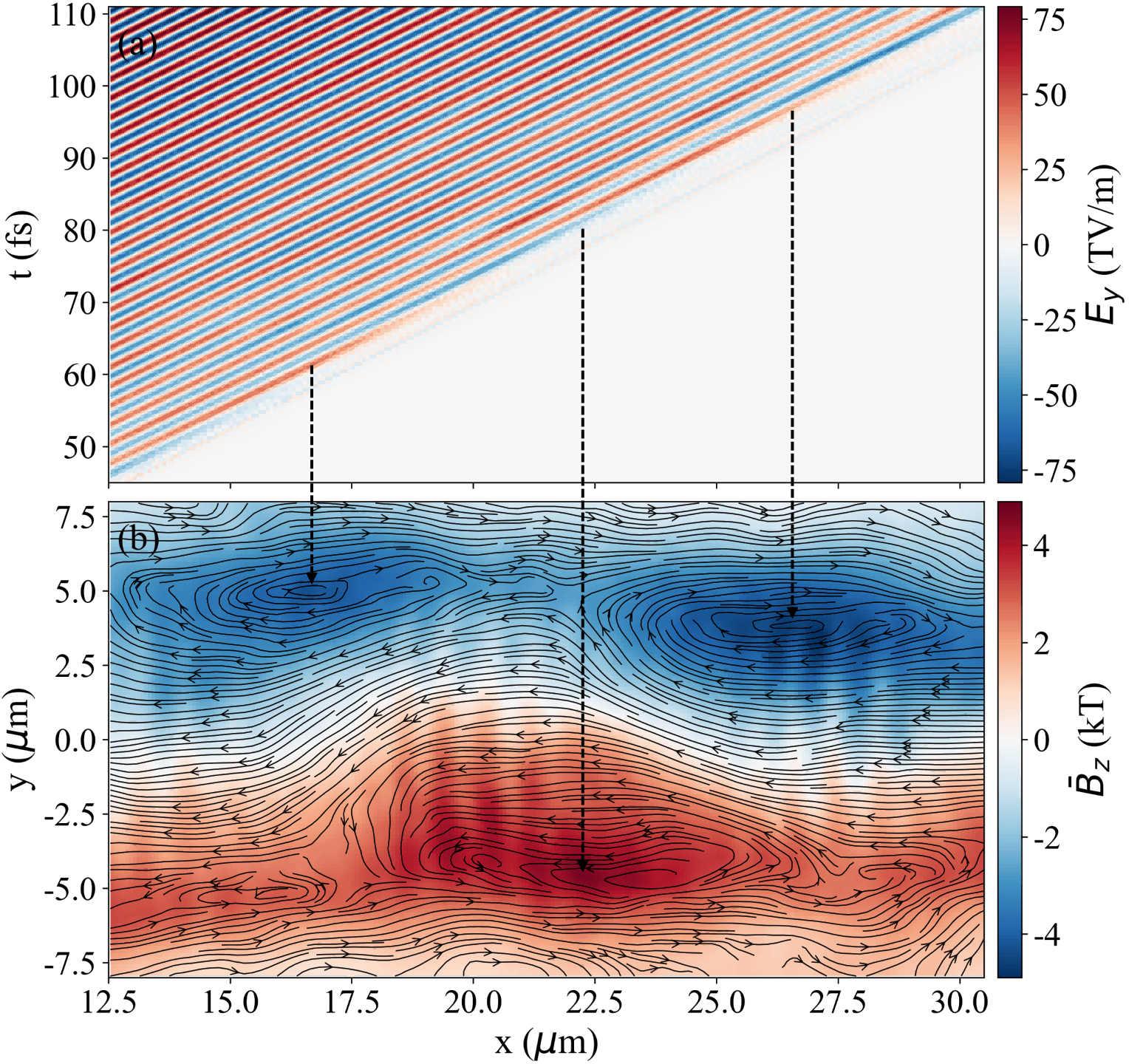}
\caption{(a) The time evolution process of the laser electric field $E_y(x)$ at the slice y = 0 when the \re{pulse} front passes through the region $x \in [10, 30] \mu m$; (b) The average azimuthal magnetic field $\bar{B}_{z}$ in the region $x \in [10, 30] \mu m$, with the arrow lines representing the average current field.
The downwards dashed arrow lines illustrate the correspondence between $E_y$ at the laser \re{pulse} front and magnetic-island structures of $\bar{B}_{z}$.
}
\label{fig:laser_etching}
\end{figure}

This behavior has been verified in previous studies\cite{Cai2025, Meir2025}. In the main PIC simulation case, we calculate the time-averaged azimuthal magnetic field $\bar{B}_{z}$ the interval from $10\,T_0$ to $390\,T_0$, where $T_0$ is the laser period, $T_0 = \lambda_0/c$. Fig.~\ref{fig:laser_etching}(b) shows $\bar{B}_{z}$ in a selected region, together with the time-averaged current field superimposed on it. The current field is represented by a two-dimensional vector field determined by the averaged longitudinal current $\bar{J}_{x}$ and the averaged transverse current $\bar{J}_{y}$.

A magnetic-island structure can be clearly observed in $\bar{B}_{z}$, and its longitudinal distribution exhibits a periodic pattern\cite{Gong2021, Cai2025, Yue2022}. The structure of the averaged current field indicates that the current distribution is responsible for the formation of the magnetic islands, while the longitudinal periodicity is primarily determined by the transverse current modulation. In Fig.~\ref{fig:laser_etching}(a), we also show the temporal evolution of the laser electric field $E_y(x)$ along the corresponding slice at y=0. This diagnostic clearly shows pulse-front etching as the laser propagates through the plasma.

According to our theoretical picture, both the \re{pulse-front} phase modulation induced by front etching and the periodic transverse current at the \re{pile-up layer} are governed by $k_{lon}$. The comparison between the averaged field structure and the \re{pulse-front} phase shown in Fig.~\ref{fig:laser_etching} verifies the consistency between these two forms of periodic modulation.
\begin{figure}
\centering
\includegraphics[width=0.48\textwidth]{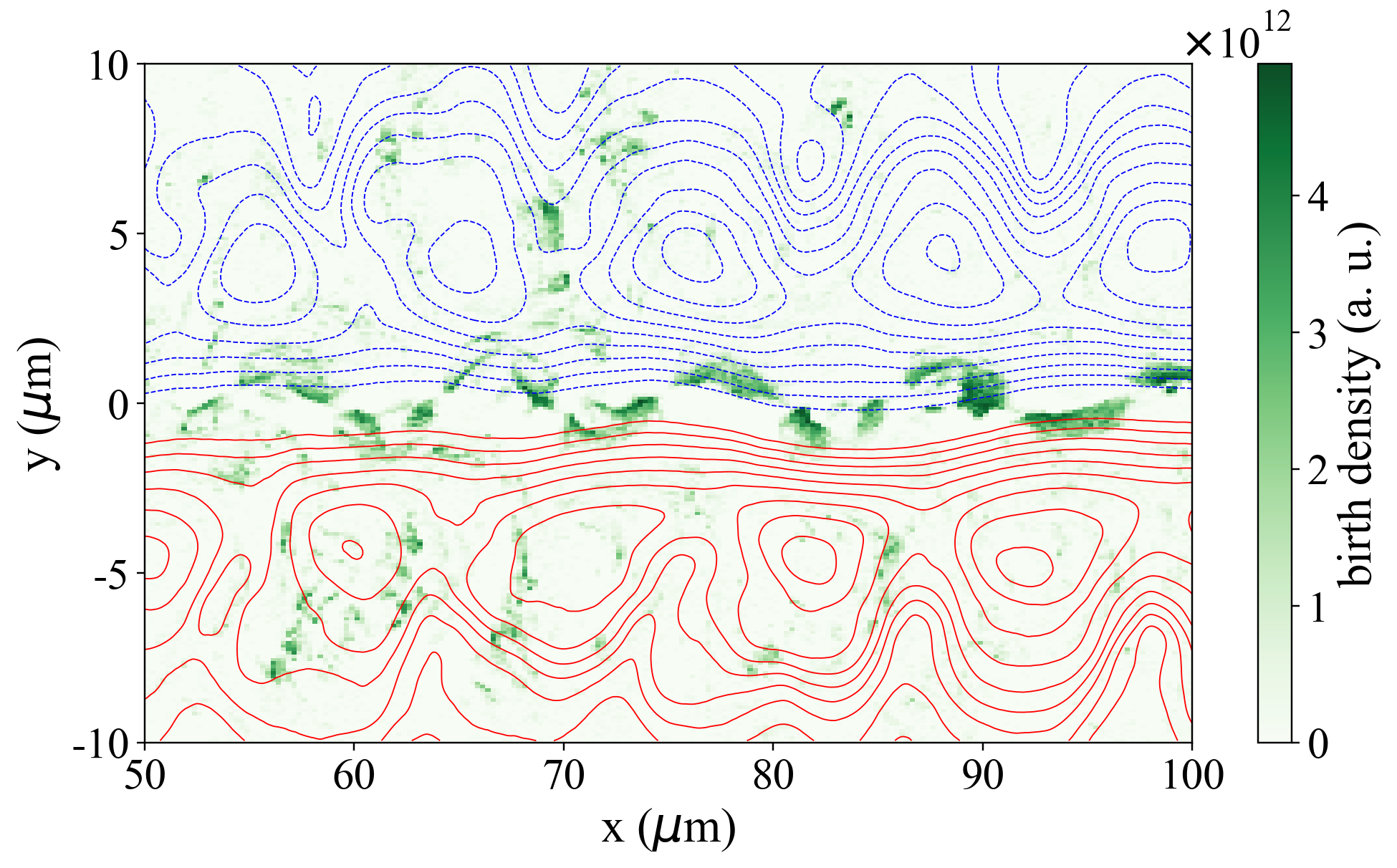}
\caption{Birth map of $x \in [50, 100] \mu m$: The distribution of the initial positions of all electrons whose energy exceeds 100 MeV at 1 ps is presented as the distribution map of initial position of injected electron. The red and blue curves represent contours of $\bar{B}_{z}$.}
\label{fig:birthmap}
\end{figure}
Furthermore, we extract the initial positions of all electrons with energies exceeding 100 MeV at 1 ps from the PIC simulation and use them as an example of the birth map of injected electrons. A local region of this birth map is shown in Fig.~\ref{fig:birthmap}. A periodic wavy structure can be clearly observed. This structure corresponds precisely to the center ridge discussed above. When the energy threshold used to select high-energy electrons is varied, the periodic wavy center ridge remains essentially unchanged. This indicates that, during direct laser acceleration, the injectable positions themselves possess a periodic geometrical structure. Such a periodic structure is predicted by our theoretical model through the center position $y_c\left(x\right)$.

We further note that the birth map shows that the initial positions of the injected electrons are mainly concentrated along the wavy center ridge. This provides additional evidence that the wavefront-guided injection mechanism plays a dominant role in supplying electrons for DLA in the laser-driven plasma channel. In Fig.~\ref{fig:birthmap}, only a local region of the birth map is shown because similar structures repeat over a longer distance. As long as the laser pulse is not depleted, the center ridge in the birth map extends farther as the laser propagates, or equivalently as the simulation duration increases. This demonstrates that the continuous injection of DLA electrons is realized through the wavefront-guided electron injection mechanism.

In laser-electron interactions in vacuum, there exists a vacuum acceleration limit\cite{Arefiev2012}. For example, for a linearly polarized plane wave, the maximum kinetic energy that an electron can gain from the laser field does not exceed $\gamma_{vac}=1+{a_0^2}/{2}$. Therefore, electron acceleration beyond the vacuum limit through direct interaction with the laser field is a characteristic feature of DLA \cite{Arefiev2012}. We also compare the birth maps obtained using different energy thresholds above $\gamma_{vac}$. The structure of the center ridge remains essentially unchanged, while only its intensity varies. This confirms that constructing the birth map by selecting high-energy electrons with an energy threshold is a reasonable diagnostic.

Next, we use the PIC simulation results to demonstrate the correlation between the position of the center ridge and the spatial phase factor $\cos(k_{lon}x+\phi_0)$. According to the theoretical model developed above, the center ridge in the birth map and the periodic spatial modulation of the averaged azimuthal magnetic field shown in Fig.~\ref{fig:laser_etching}(b) have the same origin. We therefore further overlay the spatial distribution of the averaged azimuthal magnetic field $\bar{B}_{z}$ on the birth map in Fig.~\ref{fig:birthmap}. The red and blue curves in Fig.~\ref{fig:birthmap} represent contours of $\bar{B}_{z}$. It can be clearly seen that the stripe-like injectable regions of electrons are consistent with the longitudinally periodic distribution of $\bar{B}_{z}$. This correspondence reflects the fact that, in our theoretical model, $y_c(x)$ and $\bar{B}_{z, lon} (x)$ share the same spatial phase factor, $k_{lon}x+\phi_0$.

Taken together, the PIC simulation results indicate that, due to the wavefront-guided electron injection mechanism, a center ridge with the spatial wavelength $\lambda_{lon}$ forms in the region where direct laser acceleration is driven by the ultraintense ultrashort laser. This center ridge represents the wavy stripe-like region from which electrons can be injected.
To further confirm the connection between laser-wavefront modulation and the birth map, we performed a parameter scan. We first define the ridge wavelength in the birth map as $\lambda_R$. Using the temporal evolution of the laser electric field $E_y(x)$, as shown in Fig.~\ref{fig:laser_etching}(a), we can extract the etching-dominated laser-front velocity directly from the time evolution of the electric-field distribution in the PIC simulations. This allows us to determine $\lambda_{lon}$ from the laser evolution itself.

We then compare the modulation wavelength $\lambda_{lon}$ obtained from the laser wavefront with the ridge wavelength $\lambda_R$ measured from the birth map. The agreement, $\lambda_{lon}\simeq\lambda_R$, confirms that the birth-map structure and the laser-wavefront modulation share the same physical origin. This result provides further evidence for the wavefront-guided electron injection mechanism.
\begin{figure}
\centering
\includegraphics[width=0.48\textwidth]{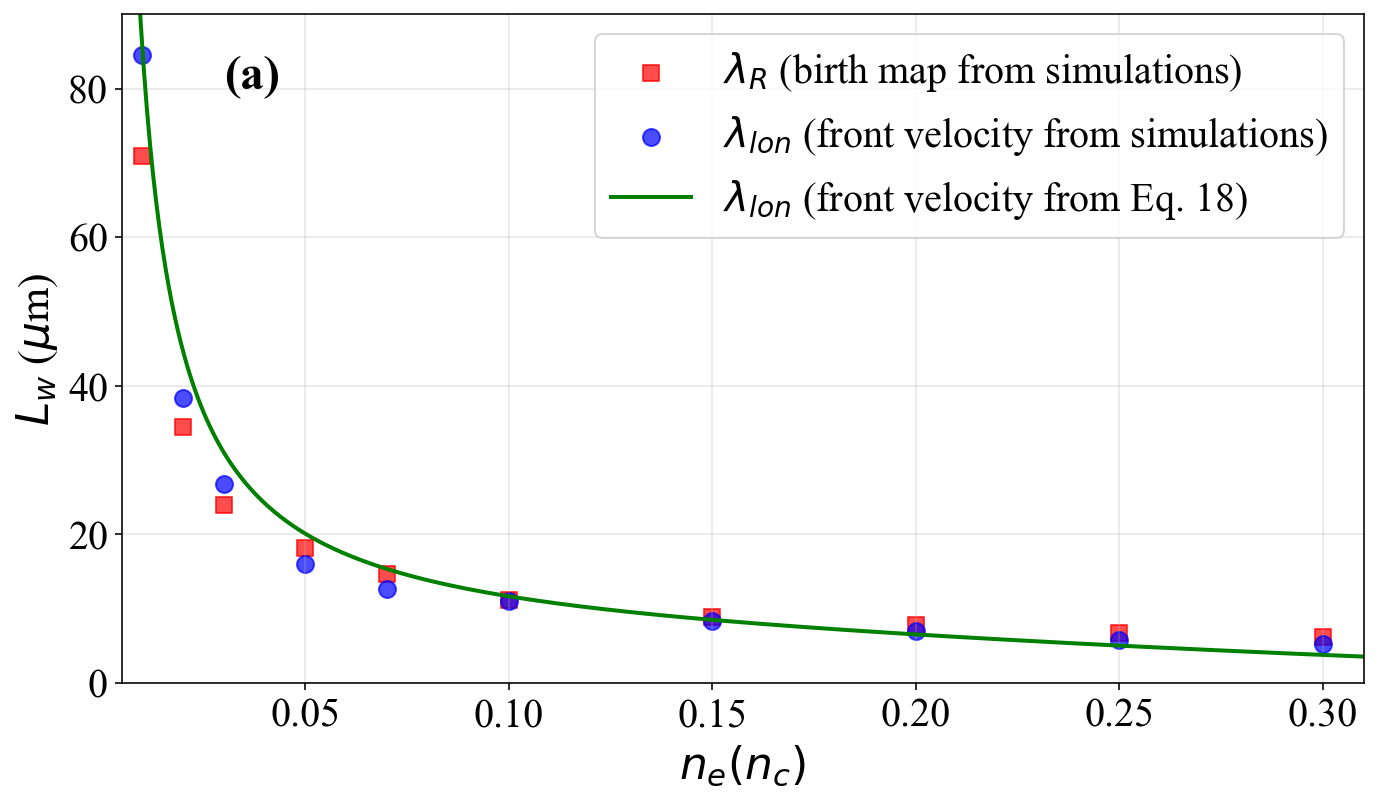}
\centering
\includegraphics[width=0.48\textwidth]{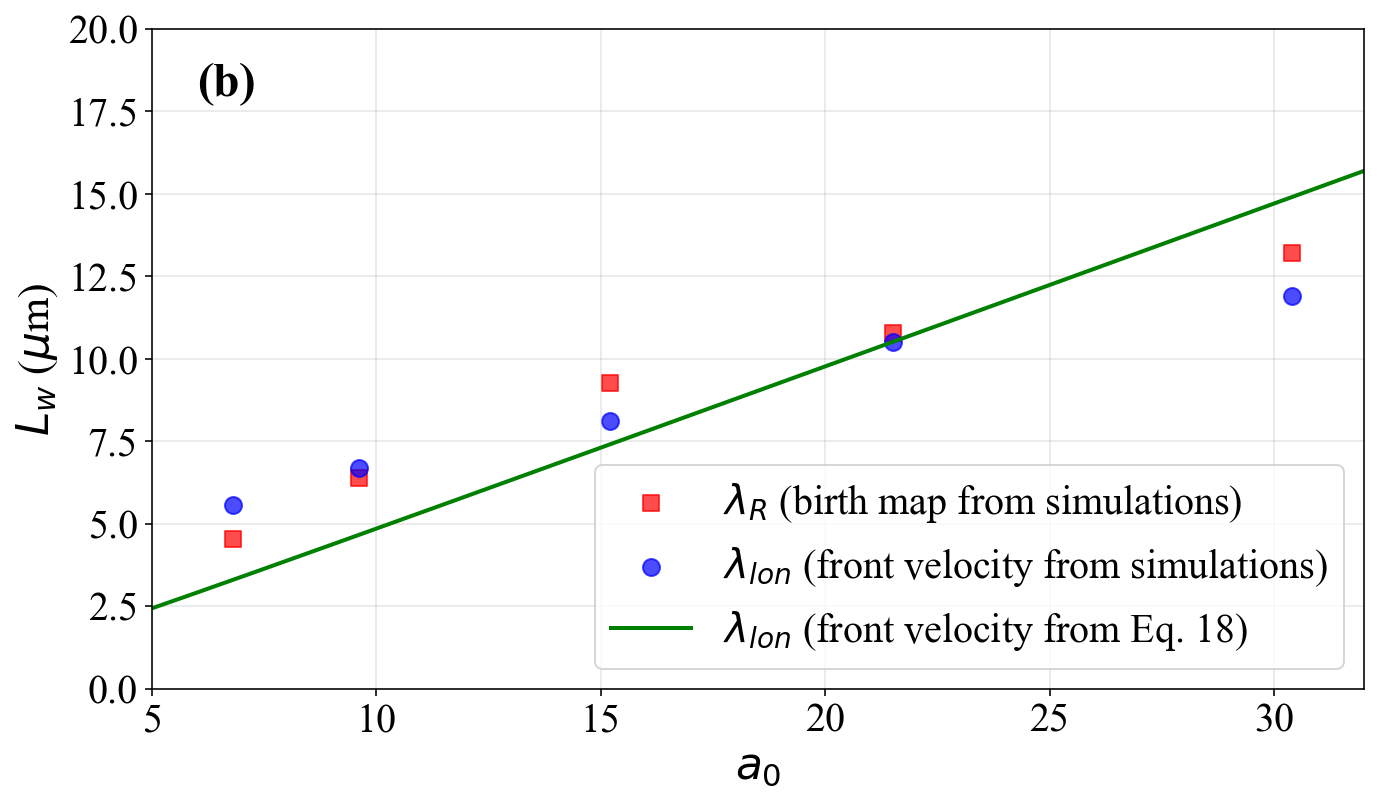}
\caption{(a) At different plasma densities, the spatial wavelength of the wavefront modulation $\lambda_{lon}$ calculated based on the laser front velocity and the spatial wavelength of the central ridge $\lambda_R$, with the normalized field amplitude $a_0 = 21.5$. (b) At different laser intensities, the spatial wavelength of the wavefront modulation $\lambda_{lon}$ calculated based on the laser front velocity and the spatial wavelength of the central ridge $\lambda_R$, with the plasma density $n_0 = 2 \times 10^{20} cm^{-3}$.}
\label{fig:para_scan}
\end{figure}
Next, we compare the spatial period of the wavefront modulation with the spatial period of the center stripe in the initial-position distribution of injected electrons. To verify this relation, we performed PIC simulations with different simulation parameters and extracted these two quantities from the results. We first varied the plasma density while keeping the other parameters fixed. The density range was extended beyond the near-critical-density regime and included $0.01n_c$, $0.02n_c$, $0.03n_c$, $0.05n_c$, $0.07n_c$, $0.1n_c$, $0.15n_c$, $0.2n_c$, $0.25n_c$, and $0.3n_c$. As shown in Fig.~\ref{fig:para_scan}(a), the two sets of data agree well in both magnitude and variation trend.

We then varied the laser intensity while keeping the other parameters unchanged, with $a_0=6.8$, $a_0=9.6$, $a_0=15.2$, $a_0=21.5$, and $a_0=30.4$. As shown in Fig.~\ref{fig:para_scan}(b), the spatial period of the wavefront modulation remains consistent with the spatial period of the center stripe of the injectable positions. Based on the parameter-scan results from the PIC simulations and the analysis of the electron-injection dynamics, we confirm that the wavefront-guided electron injection mechanism is a general feature of the direct laser acceleration process.

\begin{figure*}
\centering
\includegraphics[width=0.8\textwidth]{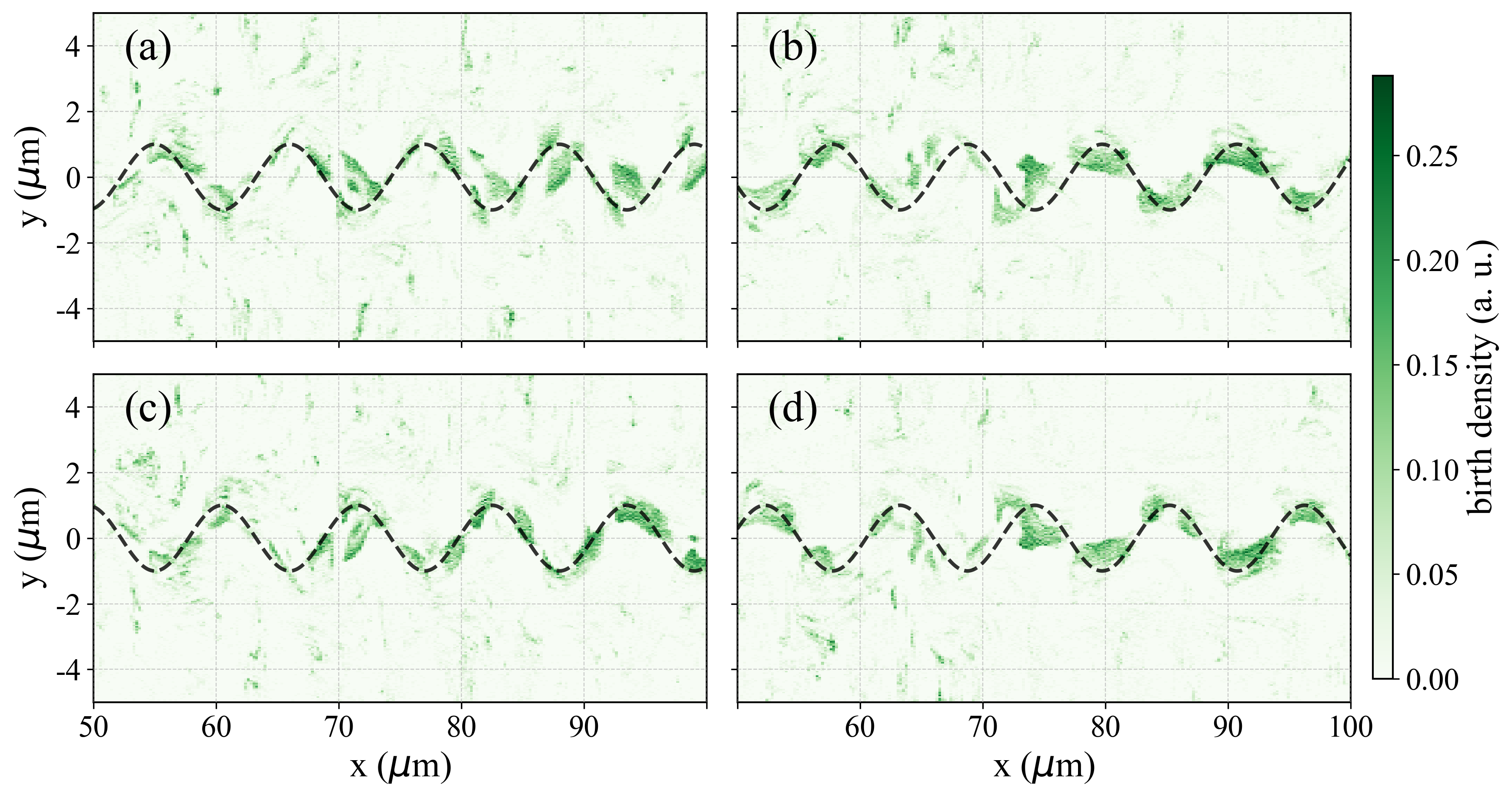}
\caption{The birth map of injected electrons with energy exceeding 100 MeV at 1 ps, corresponding to different initial phases of laser: (a) $\phi_0=0$, (b) $\phi_0 = \pi/2$, (c) $\phi_0 = \pi$, (d) $\phi_0 = 3\pi/2$. The black dashed lines in the plot represent the waveform of the function $\cos[(2\pi/\lambda_{\rm lon})x-\phi_0]$, where $\lambda_{\rm lon}=11.0 \mu m$.}
\label{fig:birthmap_phase}
\end{figure*}

Finally, we examine the full wavefront-modulation factor, $\cos(k_{lon}x+\phi_0)$. It can be seen that the periodic structure is determined not only by $k_{lon}$, but also by the phase $\phi_0$. According to our theoretical model, $\phi_0$ corresponds to the initial phase of the laser. We therefore performed PIC simulations with different initial laser phases and extracted the corresponding birth maps of the injected electrons. The four birth maps shown in Fig.~\ref{fig:birthmap_phase} correspond to otherwise identical simulation parameters, with $\phi_0=0$, $\pi/2$, $\pi$, and $3\pi/2$, respectively. By comparing Figs.~\ref{fig:birthmap_phase}(a)-~\ref{fig:birthmap_phase}(d), we obtain two conclusions. First, all cases with different $\phi_0$ exhibit the periodic feature of wavefront-modulated electron injection, and the spatial period remains unchanged. Second, the phase of the center ridge shifts with $\phi_0$. In particular, when $\phi_0$ differs by $\pi$, the center ridge is flipped across the y=0 axis. This second observation directly reflects the effect of the laser phase $\phi_0$, demonstrating its phase-shifting influence on $y_c(x)$ through the wavefront-modulation factor $\cos(k_{lon}x+\phi_0)$. 

Through the above analysis, we establish a direct correspondence among the transverse current modulation at the \re{laser-pulse} front, the longitudinally periodic quasistatic magnetic-field structure, and the birth-map distribution of DLA-injected electrons. This correspondence shows that the wavy injection ridge is a phase-locked structure imposed by the \re{pulse front}. The agreement among these independent diagnostics provides support for the proposed wavefront-guided injection mechanism and demonstrates that the initial positions of DLA electrons are organized by the same modulation process that produces the quasistatic magnetic-island structure in the plasma channel.

\section{discussion}

% The results presented above show that electron injection in DLA is not a purely random process. Instead, it is dynamically organized by the modulation of the laser front and by the local field structure of the laser-driven plasma channel. In particular, the initial positions of the electrons that are eventually injected exhibit a pronounced stripe-like periodic pattern in the birth map. This feature indicates that the electrons entering the DLA phase are not sampled indiscriminately from the background plasma. Rather, they undergo a spatially selective injection process near the laser front. 
% In this sense, the laser front is not merely the leading boundary of the propagating pulse. It acts as a dynamical injection interface that periodically modulates the local electron density and quasistatic field distribution, thereby forming spatial windows for injectable electrons. The DLA process itself relies on the coupling, or resonance, between the transverse electron oscillation and the Doppler-shifted laser field, a mechanism that has been extensively discussed in relativistic laser-channel studies \cite{Pukhov1999,Gahn1999,Arefiev2012,Robinson2013}. The present results further show that, before electrons enter the stage of sustained energy gain, the laser-front modulation has already imposed a spatial selection on the electron population.

\re{The results presented above demonstrate that electron injection into direct laser acceleration is a dynamically organized process governed jointly by the local optical-phase structure within the evolving laser-pulse front and the associated field structure of the plasma channel. We refer to this process as wavefront-guided injection. The term emphasizes that, as the pulse front propagates and evolves, successive optical wavefronts intersect the pulse-front region, causing electrons accumulated in the front-side density pile-up layer to sample different local phases of the laser field. Together with the accompanying quasistatic fields, these phase-resolved laser fields impart different transverse impulses to the electrons and thereby select those that can leave the pile-up layer and enter the channel. The laser-pulse front, the electron-density pile-up layer, and the localized channel fields thus form a self-consistent, dynamically evolving injection interface. Within this picture, the periodic ridges in the birth map represent the statistical imprint of the phase-selective injection process on the initial positions of the injected electrons.}

% First, a relativistically intense laser pulse propagating in plasma undergoes various nonlinear evolution processes. Relativistic self-focusing, pulse-front erosion, local pump depletion, self-steepening, and plasma-density modulation can all modify the local waveform of the laser pulse. As a result, the laser front does not remain an ideal, smooth, and stationary propagation boundary \cite{Decker1996,Streeter2018}. The electromagnetic field experienced by electrons near the front therefore has a strong spatiotemporal locality. Even if two electrons are initially located in nearby regions, they may experience different transverse deflections and return dynamics because they are located at different phases of the modulated front, different positions of the density pile-up, or different local field structures. The periodic stripes in the birth map can thus be understood as the projection of this nonlinear front modulation onto the initial positions of injectable electrons. In other words, the observed periodicity is not an externally imposed structure, but a spatial selection effect self-consistently generated by the evolving laser front in plasma.

\re{Wavefront-guided injection describes the self-consistent process through which electrons enter the DLA region. In a strongly nonlinear laser–plasma interaction, individual electrons may be perturbed by transient field structures, local density fluctuations, channel-boundary inhomogeneities, or the collective fields of the accelerated electron population, and their trajectories may therefore deviate from the dominant injection pathway. Nevertheless, the statistical distribution in the birth map shows that the dominant injected population follows a common physical sequence. The central contribution of this work is thus not to reproduce every microscopic detail of each individual trajectory, but to identify the dominant physical mechanism that organizes the injection of electrons within a complex nonlinear interaction. And the mechanism explains the origin of this periodic structure and reveals how the intrinsic evolution of the laser front continuously supplies electrons to the DLA region during propagation.}

\re{Another issue worth discussing is the temporal resolution used in the simulation. Insufficient temporal resolution can introduce errors in relativistic electron motion through inaccurate particle pushing near low-$\gamma$, high-field regions and through temporal interpolation of the electromagnetic fields. These local errors can perturb the electron–laser dephasing and become progressively amplified during prolonged, phase-sensitive DLA, thereby affecting the calculated trajectory and energy gain\cite{Arefiev2015Temporal, Tangtartharakul2021Integrator}. However, in terms of the correlation between the electron injection process and the front phase, the current simulation accuracy is already sufficient. We also confirmed the numerical convergence through more detailed simulations with higher resolution.}

\re{The numerical analysis in this study is based primarily on two-dimensional, three-velocity-component particle-in-cell simulations. This geometry retains all three particle-momentum components while resolving the longitudinal and in-plane transverse dynamics of electrons driven by a linearly polarized laser. It therefore captures the principal phase-dependent transverse motion, front-side density accumulation, and particle–field correlations required to identify the injection pathway considered here. Nevertheless, translational invariance in the out-of-plane direction suppresses out-of-plane electron motion and excludes finite three-dimensional focusing, cylindrical or azimuthally varying channel-field structures, transverse particle loss, and genuinely three-dimensional instabilities. These effects may modify the pulse-front evolution, local phase structure, channel fields, electron dephasing, injection probability, and subsequent energy gain. Quantitative quantities such as the injection efficiency, ridge contrast, and electron-energy distribution may therefore differ in a fully three-dimensional geometry. The present two-dimensional simulations should accordingly be regarded as establishing the physical mechanism and its dominant in-plane dynamics, rather than as providing fully quantitative predictions for a three-dimensional experimental configuration. The essential ingredients identified here are not intrinsically restricted to two dimensions, but their robustness and quantitative importance in three dimensions require further verification.}

\section{conclusion}
In summary, we have proposed a wavefront-guided electron injection mechanism for direct laser acceleration. We demonstrate for the first time that the initial positions of all injectable electrons exhibit a pronounced periodic structure, and this feature is identified as an important signature of DLA electron injection in laser-driven plasma channels. This mechanism links laser-front evolution, local quasistatic field structures, transverse electron selection, and subsequent DLA energy gain, thereby providing a unified physical picture for sustained electron injection in DLA. 
Our results show that, even in a strongly nonlinear laser-plasma interaction with complex local perturbations and individual nonideal trajectories, a dominant mechanism controlling electron injection can still exist. 
This understanding could enrich the physical picture of DLA as an efficient but intrinsically complex energy-transfer process. It also potentially provides a basis for further controlling high-charge electron beams and for developing advanced compact particle accelerators and radiation sources for high-energy photons based on relativistic laser-plasma interactions.

\begin{acknowledgments}
This work is support by the Strategic Priority Research Program of the Chinese Academy of Sciences (Grant No. XDB1550100).
The work of L. Sheng is supported by the Project for the Advanced Pulsed Radiation Imaging Technology Research Team of National Key Laboratory of Intense Pulsed Radiation Simulation and Effect. 
Z. Gong acknowledges the support from the National Key R\&D Program of China (2025YFF0515103), the CAS Project for Young Scientists in Basic Research (Grant No. YSBR-141), and the HPC Cluster of ITP-CAS for providing computational resources.
The code EPOCH is funded by the United Kingdom Engineering and Physical Sciences Research Council Grants No. EP/G054950/1, No. EP/G056803/1, No. EP/G055165/1, and No. EP/ M022463/1. 

\end{acknowledgments}

\section*{Author Declaration}
The authors have no conflicts to disclose.

\section*{Data Availability}
The data that supports the findings of this study are available from the corresponding author upon reasonable request.

\bibliography{aa}
\end{document}